\documentclass[%
 reprint,
 amsmath,amssymb,
 aps,
]{revtex4-2}
\usepackage{gensymb}
\usepackage{graphicx}% Include figure files
\usepackage{bm}% bold math

\usepackage[english]{babel}
\usepackage{float}
\usepackage{blindtext}
\usepackage{soul}
\usepackage[colorlinks=blue, urlcolor=blue, filecolor=blue]{hyperref}
\usepackage{tikz}
\usepackage{amsmath}

\usepackage{array}[=2016-10-06]  % for p columns in tabular

\begin{document}

\title{Layer-dependent Land\'e $g$-factors of electrons, holes, and excitons in two-dimensional Ruddlesden-Popper lead halide perovskites }

\author{Nataliia E. Kopteva$^{1}$, Dmitri~R.~Yakovlev$^{1}$, Mikhail~O.~Nestoklon$^{1}$, Carolin~Harkort$^{1}$, Evgeny~A.~Zhukov$^{1}$, Dennis Kudlacik$^{1}$, Erik Kirstein$^{1,2}$, Scott A. Crooker$^{2}$,  Oleh~Hordiichuk$^{3,4}$, Ole F. Dressler$^{3,4}$,  Maksym~V.~Kovalenko$^{3,4}$, and Manfred~Bayer$^{1,5}$}

\affiliation{$^1$Experimentelle Physik 2, Technische Universit\"at Dortmund, 44221 Dortmund, Germany}
\affiliation{$^{2}$National High Magnetic Field Laboratory, Los Alamos National Lab,
Los Alamos, NM 87545, USA}
\affiliation{$^{3}$Laboratory of Inorganic Chemistry, Department of Chemistry and Applied Biosciences,  ETH Z\"{u}rich, CH-8093 Z\"{u}rich, Switzerland}
\affiliation{$^{4}$Laboratory for Thin Films and Photovoltaics, Empa-Swiss Federal Laboratories for Materials Science and Technology, CH-8600 D\"{u}bendorf, Switzerland}
\affiliation{$^{5}$Research Center FEMS, Technische Universit\"at Dortmund, 44227 Dortmund, Germany}

\date{\today}

\begin{abstract}
Two-dimensional Ruddlesden-Popper lead halide perovskites provide a valuable platform for tailoring charge and spin properties through quantum confinement and reduced symmetry. While the electron and hole Landé $g$-factors in bulk lead halide perovskites exhibit a universal dependence on the band gap energy, their evolution in two-dimensional perovskites has remained largely unexplored. Here, the Zeeman splittings of electrons and holes in (PEA)$_2$MA$_{n-1}$Pb$_n$I$_{3n+1}$ perovskites with the number of inorganic layers ovarying in the range $n=1,...,8$  are measured by means of the spin-flip Raman scattering and time-resolved Kerr rotation magneto-optical techniques. A systematic evolution of the electron and hole $g$-factors with decreasing layer thickness, which deviates from the universal bulk behavior and reveals confinement-driven trends similar to those observed in perovskite nanocrystals, is found. The experimental results are in good qualitative agreement with empirical tight-binding calculations. The exciton $g$-factors are evaluated from the Zeeman splittings of the exciton resonances in reflectivity measured in pulsed magnetic fields up to 55~T.  These results provide comprehensive insight into the spin properties of two-dimensional lead halide perovskites and establish them as a tunable platform for engineering spin-dependent phenomena in quantum-confined semiconductors.

\end{abstract}

\maketitle

\begin{figure*}[t!]
\begin{center}
\includegraphics[width = 0.8\textwidth]{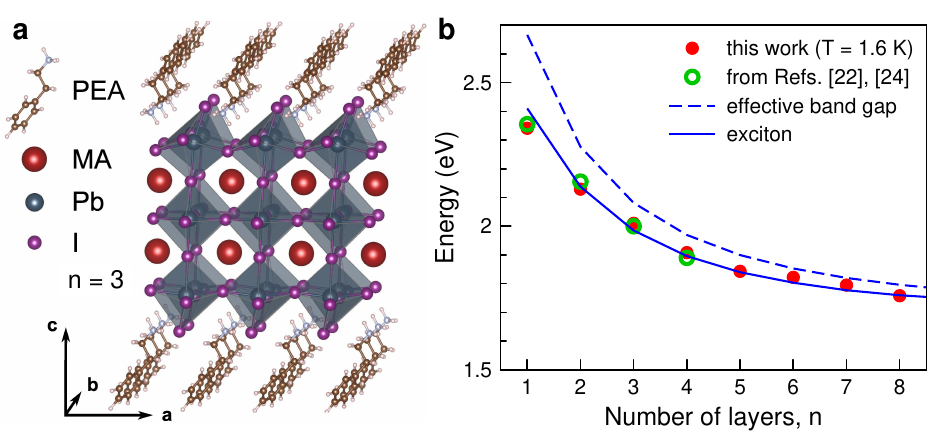}
\caption{(a) Scheme of the atom arrangement in (PEA)$_2$MA$_{2}$Pb$_3$I$_{10}$ ($n=3$). (b) Energies of the exciton resonances evaluated from reflectivity spectra of (PEA)$_2$MA$_{n-1}$Pb$_n$I$_{3n+1}$ samples as function of the number of inorganic layers $n$ at $T =1.6$~K (red symbols). The calculated effective band gap energy and exciton energy are shown by the blue dashed and solid lines, respectively. The calculated exciton binding energy is taken from Ref.~\cite{Movilla2025}, accounting for the dielectric contrast and polaron effects. The open green symbols correspond to the measured exciton energies reported in Refs.~\cite{Posmyk_2024,Dyksik_2021}. }
\label{fig:BG}
\end{center}
\end{figure*}
%\section{Introduction}

The two-dimensional (2D) lead halide perovskites have emerged as a versatile platform for exploring quantum confinement, strong spin–orbit coupling, and excitonic phenomena. In particular, the Ruddlesden-Popper perovskites combine atomically thin inorganic layers with organic spacer layers, enabling the precise control of dimensionality through the number of inorganic layers $n$. This tunability has positioned Ruddlesden-Popper perovskites as promising materials for optoelectronics and spin-based applications, including light-emitting diodes, photodetectors, THz photonic devices, and emerging spintronics concepts~\cite{Mao2019,Blancon2020,Vardeny2022_book,Vinattieri2021_book,Chu2024,Kumar2020}.  

Lead halide perovskites have recently attracted considerable attention in spin physics due to giant optical spin orientation of excitons and charge carriers~\cite{kopteva2024_FAPbI3_OO,kudlacik2024_OO_carriers}, millisecond transverse and longitudinal spin relaxation times~\cite{Meliakov2026_SML_FAPI,belykh2026_ms_T1_CsPbI3_NCs}, suppression of conventional spin relaxation mechanisms~\cite{Kopteva_2025OOX}, strong hyperfine coupling to nuclear spins~\cite{kotur2026_nucleiFAPI}, and spin coherence persisting up to room temperature~\cite{Lin_2023_RT_coherence}.  Understanding spin-dependent phenomena in these materials requires precise knowledge of the parameters governing the spin splittings.
Among them, the Land\'e $g$-factors of electrons and holes as well as excitons play a central role, as they determine the Zeeman splittings and set the energy scales relevant for the spin dynamics, hyperfine interactions, and spin manipulation. Beyond their direct relevance to spin phenomena, the $g$-factors provide valuable insight into the electronic band structure and charge-carrier properties, that are directly linked to the carrier effective masses, band arrangement, and spin-orbit interaction. 

In lead halide perovskites, where the charge-carrier mobility is low, transport-based experimental approaches have limited potential, making optical and magneto-optical techniques particularly important. In bulk lead halide perovskites, optical spectroscopy studies have revealed a universal dependence of the electron and hole $g$-factors on the band gap energy across a broad range of compositions, establishing the possibility of a unified description of the spin properties in three-dimensional systems~\cite{Kirstein_universal}. Quantum confinement, however, is expected to modify this behavior. Indeed, deviations from the bulk universal dependence have recently been demonstrated in lead halide perovskite nanocrystals (NCs)~\cite{Nestoklon_2023,Meliakov_2024_3}, highlighting the sensitivity of the $g$-factors to reduced dimensionality. 

More generally, low-dimensional materials demonstrate rich spin phenomena arising from their reduced symmetry and quantum confinement. In two-dimensional semiconductors, confinement-induced band mixing, enhanced Coulomb interaction, and symmetry breaking can strongly modify the spin level structure and the $g$-factors compared to bulk materials. While these effects have been extensively explored in van der Waals 2D materials, such as transition metal dichalcogenides~\cite{Stier_2018,Pucko_2026}, their manifestation in 2D lead halide perovskites has remained far less studied.
In Ruddlesden-Popper perovskites, the thickness of the inorganic layer provides an additional degree of freedom for tailoring the charge and spin properties. Optical and magneto-optical studies demonstrated pronounced changes in the exciton binding energies, oscillator strengths, and fine structure with varying $n$~\cite{Blancon_2018,Fang_2020,Surrente_2021,Dyksik_2021,Posmyk_2022,Posmyk_2024}. This places the 2D perovskites among the broader class of 2D semiconductors with parameters comparable to those of the van der Waals materials, including their exciton binding energies reaching several hundreds of millielectronvolts~\cite{Dyksik_2021JPCL,Movilla_2021}. Despite this progress, systematic experimental investigations of the electron and hole $g$-factor dependences on layer number, chemical composition, and organic spacer in 2D perovskites are still lacking.

Here, we investigate experimentally and theoretically the electron and hole $g$-factors in the Ruddlesden-Popper perovskites (PEA)$_2$MA$_{n-1}$Pb$_n$I$_{3n+1}$ with the number of inorganic layers varying from $n=1$ up to 8. Using high-resolution magneto-optical spectroscopy, including spin-flip Raman scattering (SFRS) and time-resolved Kerr rotation (TRKR), we measure the Zeeman splittings of electrons and holes. By applying magnetic fields in different orientations, we evaluate the components of the electron and hole $g$-factor tensors and quantify their anisotropy. The $g$-factors exhibit a systematic dependence on layer thickness, departing from the universal bulk behavior and revealing confinement-related trends previously observed in perovskite nanocrystals. This behaviour is well reproduced by empirical tight-binding calculations. In comparison with nanocrystals, a pronounced $g$-factor anisotropy is found to be specific for the 2D perovskites. We evaluate the exciton $g$-factors from the Zeeman splittings of the exciton resonances in reflectivity spectra measured in pulsed magnetic fields up to 55~T. 

\section{Results and discussion}

\subsection{Optical properties}

We study two-dimensional Ruddlesden-Popper-type lead halide perovskites of composition (PEA)$_2$MA$_{n-1}$Pb$_n$I$_{3n+1}$. These materials consist of layers of corner-sharing PbX$_6$ octahedra that form quantum wells with methylammonium (MA), which are separated by van der Waals-bonded organic spacer layers composed of phenethylammonium (PEA) as shown in Figure~\ref{fig:BG}(a) (for details see Supplementary Information \ref{sec:SI:calcul}).  The number of inorganic layers $n$ determines the thickness of the inorganic quantum well and thus provides a direct tool to control dimensionality and carrier quantum confinement.

\begin{figure*}[t!]
\begin{center}
\includegraphics[width = 0.99\textwidth]{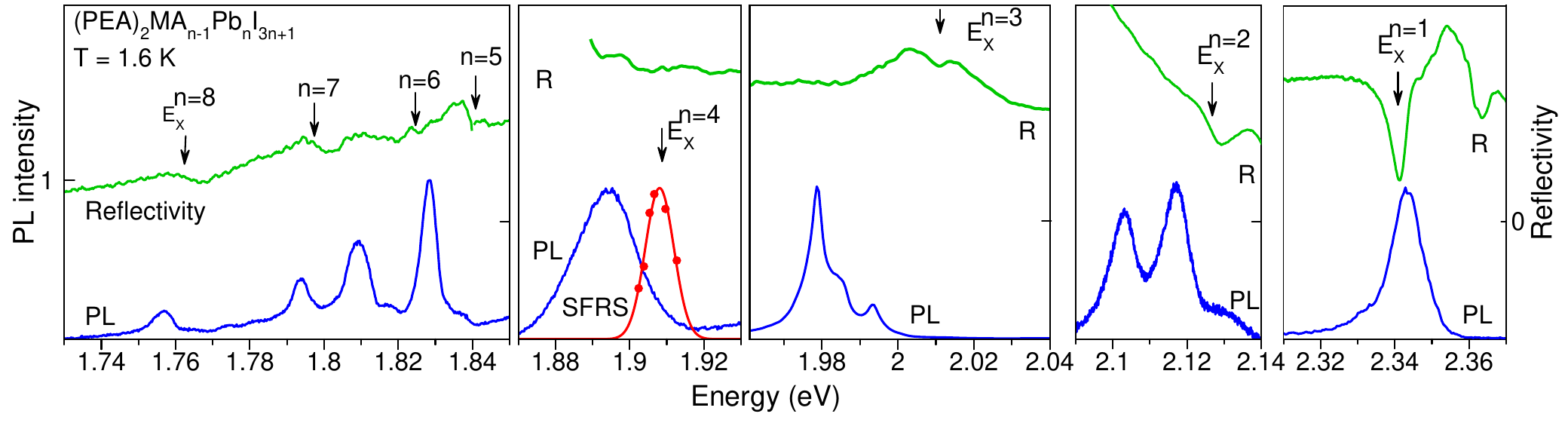}
\caption{Optical properties of 2D perovskites (PEA)$_2$MA$_{n-1}$Pb$_n$I$_{3n+1}$: photoluminescence (blue) and reflectivity (green) spectra measured at $T = 1.6$\,K. The red symbols and line show the intensity of the SFRS signals for the sample with $n=4$.
}
\label{fig:PL}
\end{center}
\end{figure*}

The optical properties are governed by strongly bound excitons and by an effective band gap that is renormalized from the bulk band gap due to quantum confinement and depends on the number of layers $n$. We study four (PEA)$_2$MA$_{n-1}$Pb$_n$I$_{3n+1}$ samples with different numbers of layers: $n = 1$, $n = 2$, $n = 3$, and $n = 4,...,8$. Figure~\ref{fig:PL} shows their photoluminescence and reflectivity spectra.  In the reflectivity spectra pronounced exciton resonances are observed, with energies ranging from $E^{n = 8}_\text{X} = 1.762$\,eV to $E^{n = 1}_\text{X} = 2.341$\,eV, see Table~\ref{tab:exciton}. Pronounced exciton features are observed for small $n$, reflecting a high oscillator strength enhanced by the quantum confinement, while the resonances gradually broaden and weaken as $n$ increases. In the large-$n$ limit, the system approaches the bulk case, namely MAPbI$_3$ crystals with a band gap energy of $E_\text{g} = 1.652$\,eV at cryogenic temperatures. According to ref.~\cite{Rahil_2022}, 2D perovskites with $n > 5$ enter the orthorhombic crystallographic phase at temperatures below 178~K, which is consistent with the orthorhombic structure of bulk MAPbI$_3$ crystals at temperatures below 160~K.

Comparing the photoluminescence spectra, a series of emission lines is observed, with energies in correspondence with the different numbers of inorganic layers $n$. For the $n = 1$ sample, the PL line coincides in energy with the exciton resonance in reflectivity indicating the absence of a Stokes shift. For $n > 1$, the PL lines are red-shifted with respect to the corresponding exciton resonances in reflectivity, revealing finite Stokes shifts (values are given in Table~\ref{tab:exciton}). However, the PL cannot be attributed exclusively to exciton emission and may also include contributions from recombination processes involving resident charge carriers. This is indicated by the recombination dynamics of the $n = 1$ sample at the PL maximum, which exhibit a fast decay of 20~ps associated with exciton radiative recombination, but also a slow decay of 340~ps~\cite{Kirstein_2023}, and even weak emission with decay times up to 40~$\mu$s~\cite{Harkort_2023}. These longer times can be related to recombination of localized and spatially separated electrons and holes, as is  typical also in the low-temperature PL of bulk lead halide perovskites~\cite{kopteva2024_FAPbI3_OO,kudlacik2024_OO_carriers}.

The assignment of the observed exciton resonances to specific layer numbers $n$ is based on established literature: ref.~\cite{Smith_2019} for low-$n$ and ref.~\cite{Soe_2019} for higher-$n$ iodide 2D perovskites. Figure~\ref{fig:BG}(b) summarizes the exciton resonance energies in reflectivity as a function of $n$. A clear, monotonic decrease of the exciton resonance energy with increasing $n$ is observed, confirming that the measured resonances are associated with distinct $n$. The exciton energies in the studied samples span over the spectral range $1.76-2.34$~eV, providing a platform for investigating spin-related properties in a wide range of effective band gaps in the visible.

\subsection{Spin-flip Raman scattering and time-resolved Kerr rotation}

We employ spin-flip Raman scattering to investigate the spin properties of the resident carrier spins in 2D perovskites.
In an external magnetic field $\mathbf{B}$, the spin sublevels of electrons (e) and holes (h) as well as excitons (X) are split by the Zeeman energy
\begin{equation}
E_{\text{Z},j} = g_{j}\mu_\text{B}B,
\label{eqn:EZ}
\end{equation}
which is proportional to the magnetic field strength and the corresponding electron, hole, or exciton Land\'e $g$-factor, $g_{j}$ with index $j = \text{e,h,X}$. $\mu_\text{B}$ is the Bohr magneton.

In the Raman scattering process, a carrier spin can flip its orientation, which requires either absorption or emission of an energy equal to $E_\text{Z,e(h)}$, depending on whether the spin transition occurs from the lower to the upper Zeeman sublevel or vice versa~\cite{hafele_chapter_1991,Kalitukha_2026}. As a result, the energy of the scattered photon differs from the incident laser photon energy by $E_\text{Z,e(h)}$.

\begin{figure*}[t!]
\begin{center}
\includegraphics[width = 1\textwidth]{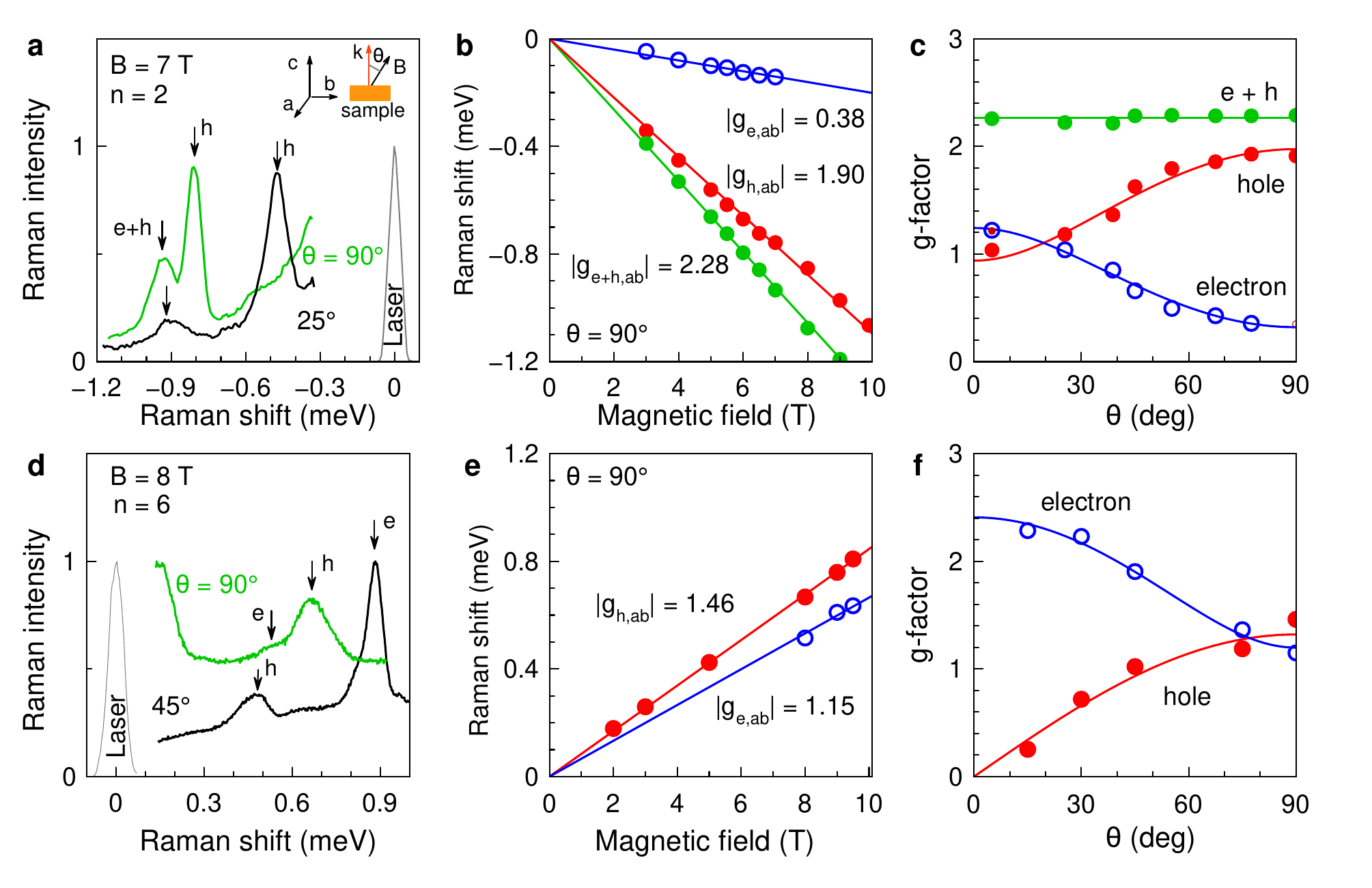}
\caption{Spin-flip Raman scattering of electrons and holes in (PEA)$_2$MAPb$_2$I$_{7}$  ($n = 2$) and (PEA)$_2$MA$_5$Pb$_6$I$_{19}$ ($n = 6$) at $T =1.6$\,K. 
(a) SFRS spectra measured at $B=7$~T applied in the Voigt geometry ($\theta = 90\degree$, green) and tilted geometry ($\theta = 25\degree$, black) on the $n = 2$ sample. $E_\text{exc} = 2.132$\,eV and $P = 0.15$\,W/cm$^2$. The spin-flips of hole (h) and combined $\rm {e+h}$ are indicated by the vertical arrows. The Sketch shows the experimental geometry.
(b) Raman shifts of hole (red symbols) and combined $\rm {e+h}$ (green symbols) flips as function of magnetic field applied in the Voigt geometry ($\theta = 90\degree$) in the $n = 2$ sample. The electron Raman shifts are calculated from the difference of combined carrier and hole Raman shifts (open blue symbols). The lines are fits with eq~\eqref{eqn:EZ} giving $g_\text{e,ab} = 0.38$, $g_\text{h,ab} = 1.90$, and $g_\text{e+h,ab} = 2.28$.  
(c) Dependence of the hole (red symbols), electron (blue symbols), and combined $\rm {e+h}$ (green symbols) $g$-factors on the angle $\theta$ in the $n = 2$ sample. Lines are fits with eq~\eqref{eqn:ani_gf}. 
(d) SFRS spectra measured at $B = 8$~T in the Voigt geometry ($\theta = 90\degree$, green) and tilted geometry ($\theta = 45\degree$, black) on the $n = 6$ sample. $E_\text{exc} = 1.814$\,eV and $P = 0.15$\,W/cm$^2$. The spin-flip transitions of electrons (e) and holes (h) are indicated by the vertical arrows.
(e) Raman shifts of holes (red symbols) and electrons (blue symbols) as function of magnetic field applied in the Voigt geometry ($\theta = 90\degree$) for the $n = 6$ sample. The lines are fits using eq~\eqref{eqn:EZ}, yielding $g_\text{e,ab} = 1.15$ and $g_\text{h,ab} = 1.46$.  
(f) Angular dependence of the hole (red symbols) and electron (blue symbols) $g$-factors in the $n = 6$ sample. The lines are fits using eq~\eqref{eqn:ani_gf}.}
\label{fig:SFRS}
\end{center}
\end{figure*}

\begin{figure*}[t!]
\begin{center}
\includegraphics[width = \textwidth]{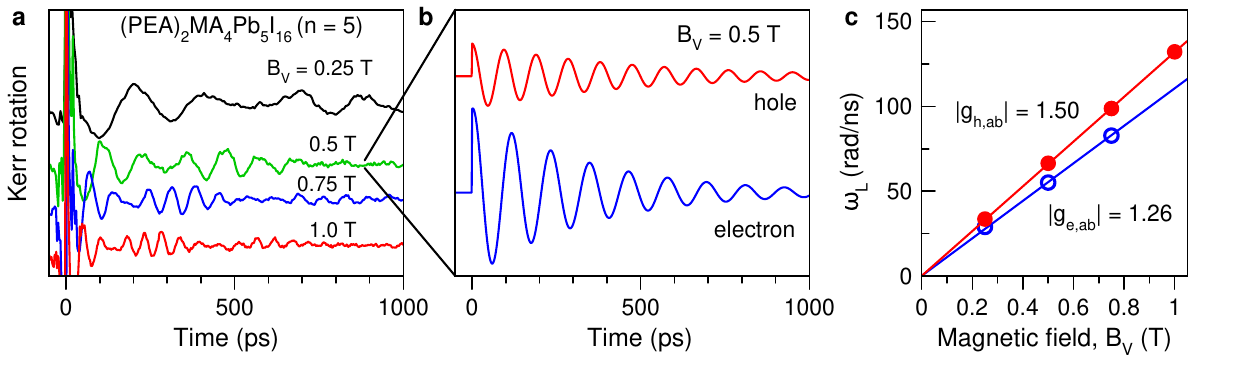}
\caption{Spin dynamics of electrons and holes in (PEA)$_2$MA$_4$Pb$_5$I$_{16}$ ($n = 5$) measured by TRKR in the Voigt geometry at $T = 1.6$\,K. 
(a) Spin dynamics in various magnetic fields. $E_\text{exc} = 1.839$\,eV. 
(b) Hole (red) and electron (blue) contributions to the TRKR signal extracted from the dynamics at $B_\text{V}= 0.5$\,T by fitting the data with eq~\eqref{eqn:KR}.
(c) Dependence of the electron (blue symbols) and hole (red symbols) Larmor precession frequencies $\omega_\text{L,(e,h)}$ on magnetic field. The lines are fits with eq~\eqref{eqn:LF} giving $|g_\text{e,ab}| = 1.26$ and $|g_\text{h,ab}| = 1.50$.}
\label{fig:TRKR}
\end{center}
\end{figure*}

In semiconductors, spin-flip Raman scattering is mediated by the exciton states, which provide efficient coupling between light and the carrier spins. This coupling is strongly enhanced when the excitation energy is tuned to the exciton resonance, owing to the large oscillator strength of the exciton optical transitions.

Figure~\ref{fig:SFRS}a shows representative SFRS spectra measured in an external magnetic field tilted with respect to the $\bf{k}$-wavevector of the light by $\theta = 25^\circ$ (see sketch in Figure~\ref{fig:SFRS}a) and $\theta = 90^\circ$ (Voigt-geometry) on the (PEA)$_2$MAPb$_2$I$_{5}$ ($n = 2$) sample at $B = 7$\,T. Distinct spin-flip Raman lines associated with the hole and the combined electron-hole spin slips can clearly be resolved, as indicated by the vertical arrows. For clarity of presentation, the hole (h) and combined carriers (e+h) labels are introduced here, with their justification discussed below.

In the Voigt geometry, where the magnetic field is applied along the 2D perovskite layers, the Raman shifts of the electron, hole, and combined electron-hole spin-flip lines increase linearly with growing magnetic field, as shown in Figure~\ref{fig:SFRS}b. Linear fits with eq~\eqref{eqn:EZ} yield $g$-factors of $|g_{\rm e,ab}| = 0.38$ for the electrons, $|g_{\rm h,ab}| = 1.90$ for the holes and $|g_{\rm e+h,c}| = 2.28$ for electron-hole combined. Figure~\ref{fig:SFRS}c summarizes the angular dependence of the electron, hole, and combined $g$-factors. The experimental data are well described using an anisotropic $g$-factor given by:
\begin{equation}
\label{eqn:ani_gf}
g(\theta)=\sqrt{(g_\text{ab}\sin \theta)^2 + (g_\text{c}\cos \theta)^2}.
\end{equation}
The electron $g$-factor reaches its maximum value in the Faraday geometry and its minimum value in the Voigt geometry, whereas the hole $g$-factor exhibits the opposite angular dependence. As a result, the sum of the electron and hole $g$-factors remains almost independent of $\theta$.

A similar SFRS behavior is observed for the thicker sample (PEA)$_2$MA$_5$Pb$_6$I$_{19}$ ($n = 6$). The SFRS spectra shown in Figure~\ref{fig:SFRS}d are  measured in tilted-field and Voigt geometries and exhibit well-resolved spin-flip lines associated with the electrons and holes. As in the $n = 2$ case, the Raman shifts of these spin-flip lines increase linearly with magnetic field, allowing for evaluation of $g_{\rm e,ab} = 1.15$ for the electrons and $g_{\rm h,ab} = 1.46$ for the holes in the Voigt geometry, see Figure~\ref{fig:SFRS}e. The angular dependence of the electron and hole $g$-factors in the $n = 6$ sample shown in Figure~\ref{fig:SFRS}f follows the anisotropic behavior described by eq~\eqref{eqn:ani_gf}. The evaluated components of the $g$-factor tensors for all studied samples are collected in Table~\ref{Table_I}.

\begin{figure*}[t]
\begin{center}
\includegraphics[width = 1\textwidth]{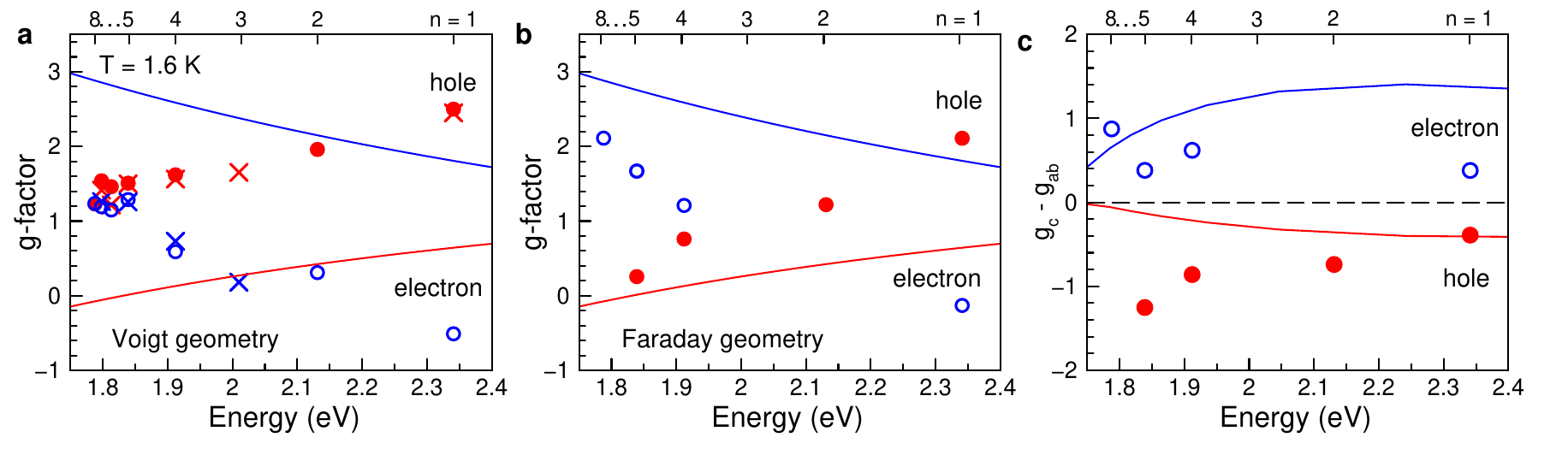}
\caption{Electron (blue symbols) and hole (red symbols) $g$-factors as a function of exciton energy in (PEA)$_2$MA$_{n-1}$Pb$_n$I$_{3n+1}$ ($n=1,...,8$) at $T = 1.6$~K. the circles are measured with SFRS and the crosses with TRKR. (a) Voigt geometry. (b) Faraday geometry. The solid lines indicate the universal dependence of the electron (blue) and hole (red) $g$-factors on the band gap energy for bulk lead halide perovskites (from ref.~\cite{Kirstein_universal}). (c) Energy dependence of the $g$-factor anisotropy, shown as the difference value $g_{\rm e(h),c}-g_{\rm e(h),ab}$. The blue and red lines show the calculation results of the electron and hole $g$-factor anisotropies in 2D perovskites as a function of the effective band gap energy, determined by the ETB method. Finite in-plane carrier localization over 19 elementary cells is considered (see Supplementary Information \ref{sec:SI:calcul}).      
}
\label{fig:MUD}
\end{center}
\end{figure*}

\begin{table*}[b!]
\caption{Summary of the electron and hole $g$-factors in (PEA)$_2$MA$_{n-1}$Pb$_n$I$_{3n+1}$ samples evaluated from SFRS and TRKR experiments. The measurement accuracy is $\pm$0.05 for SFRS and TRKR. For $n = 2,7,8$, $g_\mathrm{e,c}$ and $g_\mathrm{h,c}$ are determined by extrapolation to $\theta = 0$.}
\centering
\begin{tabular}{p{0.7cm} p{1.8cm} p{1.2cm} p{1.2cm} p{1.2cm} p{1.2cm} p{1.2cm} p{1.2cm}}
%\begin{tabular}{rrrrrrrr}
\hline
& & SFRS & SFRS & SFRS & SFRS & TRKR & TRKR\\
$n$ & $E_\mathrm{exc}$ & $g_\mathrm{e,c}$ & $g_\mathrm{h,c}$ & $g_\mathrm{e,ab}$ & $g_\mathrm{h,ab}$ & $g_\mathrm{e,ab}$ & $g_\mathrm{h,ab}$\\
\hline
1 & 2.341~eV & $-0.13$ & 2.11 & $-0.51$ & 2.50 &      & 2.45 \\
2 & 2.129~eV &  1.24    & 0.93 & 0.38 & 1.90 &  &  \\
3 & 2.010~eV & &  &  &  & 0.18 & 1.65 \\
4 & 1.908~eV & 1.21 & 0.74 & 0.59 & 1.62 & 0.73 & 1.56 \\
5 & 1.837~eV & 1.67 & 0.25 & 1.29 & 1.51 & 1.26 & 1.50 \\
6 & 1.813~eV & 2.4 & 0  & 1.15 & 1.46 &      & 1.22 \\
7 & 1.795~eV &  2.06    &   0   & 1.54 & 1.19 & 1.41 & 1.26 \\
8 & 1.762~eV & 2.11 &      & 1.23 &      &      &      \\
\hline
\end{tabular}
\label{Table_I}
\end{table*}

The relative intensities of the spin-flip Raman lines exhibit a systematic dependence on the layer thickness. For thin samples with $n < 4$, the spin-flip features are more pronounced in the anti-Stokes part of the spectrum. In contrast, upon approaching the bulk limit, the spin-flip Raman lines become increasingly dominant in the Stokes range. Such a crossover from anti-Stokes- to Stokes-dominated spin-flip scattering is in good agreement with the previously reported results for two-dimensional perovskites with $n = 1$~\cite{Harkort_2023} and for bulk CsPbBr$_3$~\cite{Kalitukha_2026}. The experimental results on SFRS in the samples with $n=4,5,7$ are given in Figures~\ref{fig:SI_n3}, \ref{fig:SI_n4}, and \ref{fig:SI_n6}.

The spin dynamics of resident carriers in 2D perovskites are also investigated by time-resolved Kerr rotation in the Voigt geometry at $T = 1.6$~K. Figure~\ref{fig:TRKR}a shows representative TRKR signals measured on (PEA)$_2$MA$_4$Pb$_5$I$_{16}$ ($n = 5$) under resonant excitation at $E_\text{exc} = 1.839$~eV in different magnetic fields. The oscillatory Kerr rotation signals originate from the Larmor precession of the electron and hole spins in the external magnetic field:
\begin{equation}
\label{eqn:LF}
\omega_\text{L,e(h)} = |g_{\rm e(h),ab}| \mu_\text{B}  B_\text{V}/\hbar.
\end{equation}
Here $\mu_\text{B}$ is the Bohr magneton, $g_{\rm e(h),ab}$ is the electron (hole) $g$-factor, $\hbar$ is the reduced Planck constant. The extracted Larmor precession frequency $\omega_\text{L}$ increases linearly with magnetic field, as shown in Figure~\ref{fig:TRKR}c. A linear fit using eq~\eqref{eqn:LF} yields  $g_{\rm e,ab} = 1.26$ and $g_{\rm h,ab} = 1.50$. We measure the carrier spin dynamics by TRKR in all samples, except for $n = 2$ and 8 where the signal was too weak, and the evaluated electron and hole $g$-factors are given in Table~\ref{Table_I}. One can see that they are in good agreement with the SFRS data so that these two techniques complement each other. 

It is important to note that the TRKR technique primarily probes the spin dynamics of the resident charge carriers, whereas the SFRS technique is sensitive to spin-flip processes involving both the excitons and resident electrons or holes. Therefore, a direct comparison of the results obtained by these two methods allows for an unambiguous identification of the exciton contribution to the observed spin signals. The assignment of the electron and hole spin resonances is further supported by their dependence on the effective band gap energy varying with the number of layers $n$. Note that in our previous studies of (PEA)$_2$PbI$_{4}$ ($n = 1$), the strong renormalization of the $g$-factors by confinement was not yet evident, which brought us to an opposite identification of the electron and hole $g$-factors~\cite{Harkort_2023,Kirstein_2023}. 

\subsection{Layer dependence of electron and hole $g$-factors}
 \begin{figure*}[hbt]
\begin{center}
\includegraphics[width = 0.8\textwidth]{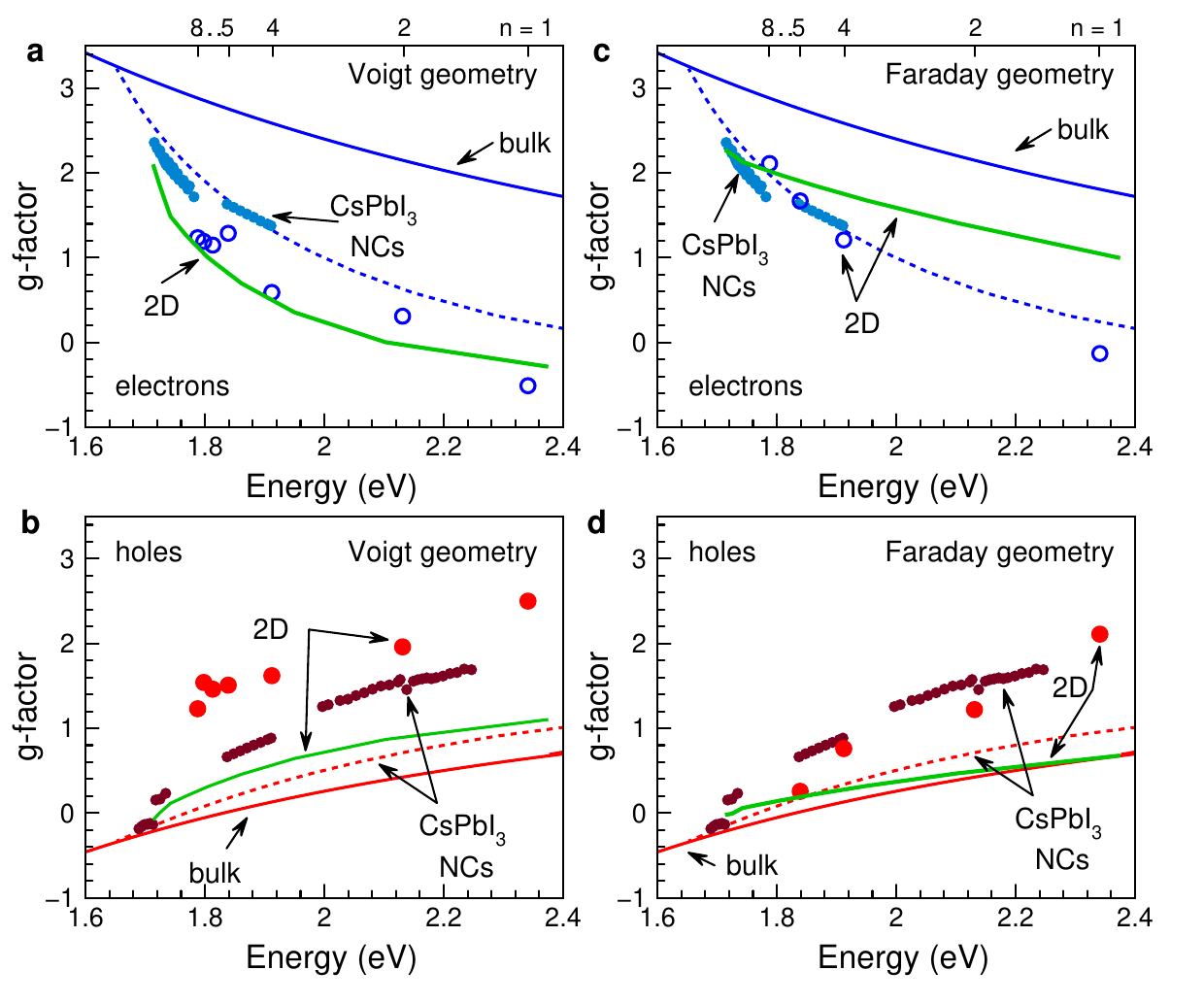}
\caption{Universal dependence of the electron and hole Land\'e $g$-factors on the band gap energy:  
%Comparison of charge carrier $g$-factors in 2D (Faraday geometry) and nanocrystals.
electron $g$-factor energy dependence measured in the Voigt (a) and Faraday (c) geometries with experimental data for the 2D samples (open symbols) and for CsPbI$_3$ NCs (closed symbols, Ref.~\cite{Meliakov_2024_3}). 
Hole $g$-factor energy dependence measured in the Voigt (b) and Faraday (d) geometries with xperimental data for the 2D (red symbols) and for CsPbI$_3$ NCs (brown symbols, Ref.~\cite{Meliakov_2024_3}).
In both panels, the lines are results of model calculations. The solid blue and red lines are for bulk lead halide perovskites from Ref.~\cite{Kirstein_universal}. The dashed lines are for CsPbI$_3$ NCs from Ref.~\cite{Nestoklon_2023}. The green lines represent the theoretical calculation of the electron and hole $g$-factors in 2D perovskites as function of the exciton energy, determined by the ETB method. Finite in-plane carrier localization over 19 elementary cells is considered (see Supplementary Information \ref{sec:SI:calcul} and Figure~\ref{fig:gETB2}(b)).
}
\label{fig:CUD}
\end{center}
\end{figure*}

Figure~\ref{fig:MUD}a summarizes the electron and hole $g$-factors measured in the Voigt geometry on the (PEA)$_2$MA$_{n-1}$Pb$_n$I$_{3n+1}$ samples. They are presented as function of the exciton energy, which reflects the effective band gap tuned by the number of inorganic layers $n$. The experimental data obtained by SFRS (circles) and TRKR (crosses) show good agreement with each other. The electron and hole $g$-factors exhibit clear and systematic dependences on $n$, in both the Voigt and Faraday geometries. The electron $g$-factor shows a systematic decrease with increasing energy (i.e. with decreasing $n$), whereas the hole $g$-factor shows the opposite trend and increases monotonically with increasing energy. 

 For comparison, the solid lines in Figures~\ref{fig:MUD}a,b represent the universal dependence of the electron and hole $g$-factors on the band gap energy established for bulk lead halide perovskites~\cite{Kirstein_universal}. One can see that in the 2D perovskites the same trends for the energy dependences of $g$-factors are valid. Namely, the electron $g$-factor decreases with increasing effective band gap, while the hole $g$-factor increases. We use this fact for the assignment of the electron and hole $g$-factors, similar to the approach we used previously for perovskite NCs~\cite{Nestoklon_2023,Meliakov_2024_3}. This correspondence indicates that, despite the reduced dimensionality and the pronounced quantum confinement, the spin properties of charge carriers in layered perovskites retain a clear connection to the band structure. However, one can see that despite the similar trends for the 2D and bulk perovskite $g$-factors, their values differ significantly. Thus the electron and hole $g$-factors are strongly renormalized by quantum confinement, in line with our previous findings for perovskite NCs~\cite{Nestoklon_2023,Meliakov_2024_3}. The reasons for the renormalization are the same in 2D and NCs, while the absolute values and the degree of anisotropy differ. We will consider that below by comparing the experimental data with the results of calculations for 2D perovskites.

It is important to note that the SFRS and TRKR techniques provide access to the value of the charge carrier $g$-factors, but not to their sign. We draw conclusions about the signs of the electron and hole $g$-factors in Figures~\ref{fig:MUD}a,b from the fact that the exciton $g$-factor is positive and has a value of about $+2$, as measured by magneto-reflectivity in high magnetic fields (see Figure~\ref{fig:X}). Here, we use the relation for a bright exciton in perovskite semiconductors: $g_{\rm X}=g_{\rm e}+g_{\rm h}$. 

By comparing the experimental results in Figures~\ref{fig:MUD}a,b one can see a considerable difference in the $g$-factor values measured in Voigt and Faraday geometry. In fact, the $g$-factor anisotropy is inherent for 2D materials and thus is expected also in 2D perovskites. We present it in Figure~\ref{fig:MUD}c as the difference $g_{\rm c} - g_{\rm ab}$, plotted as function of energy. One can see that for the electrons  $g_{\rm c} > g_{\rm ab}$ and the anisotropy moderately decreases for larger energies. The holes exhibit the opposite relation with $g_{\rm c} < g_{\rm ab}$, while the absolute value of the anisotropy  $|g_{\rm c} - g_{\rm ab}|$ also decreases for larger energies. Notably, this behavior is counterintuitive, as a stronger anisotropy is typically expected in the strongly confined regime for 2D structures. In conventional semiconductor quantum wells, the confinement-induced mixing of the light-hole and heavy-hole states leads to the pronounced $g$-factor anisotropy~\cite{bookIvchenko}.

\begin{figure*}[t!]
\begin{center}
\includegraphics[width = 0.8\textwidth]{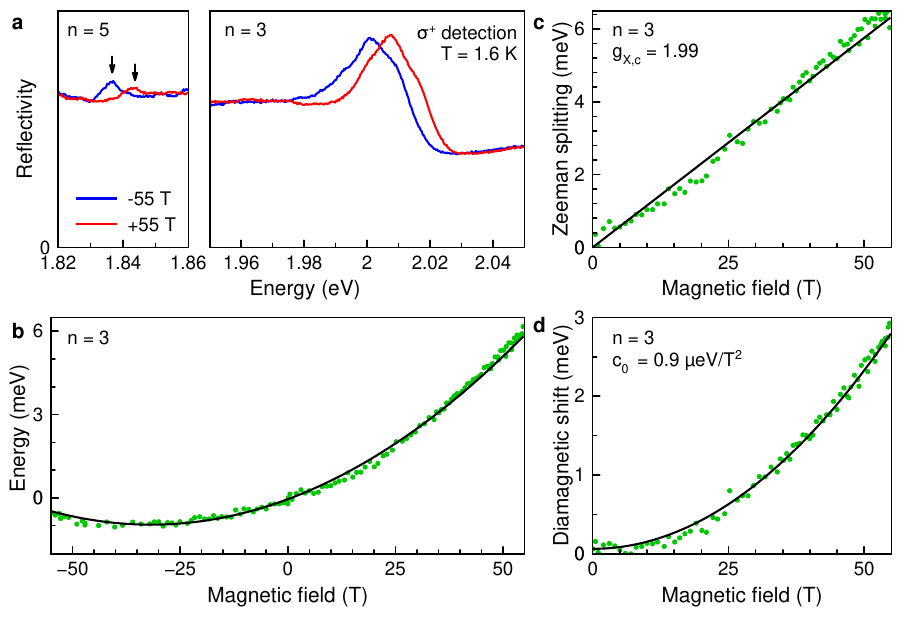}
\caption{Exciton reflectivity spectra, spin splitting, and diamagnetic shift in magnetic fields up to 55~T at $T = 1.6$~K for $(\mathrm{PEA})_2\mathrm{MA}_{n-1}\mathrm{Pb}_n\mathrm{I}_{3n+1}$. (a) Reflectivity spectrum measured in $\sigma^+$ polarization at $B_\mathrm{F} = +55$~T (red) and $B_\mathrm{F} = -55$~T (blue) for $n = 3$ and $n = 5$. Magnetic-field dependences of the exciton energy shift (b), the Zeeman splitting (c) and the diamagnetic shift (d) for $n = 3$, shown by the green symbols. The black line in panel (b) represents a fit using eq~\eqref{eq:DS}, where the first term describes the Zeeman splitting (black line in panel (c)), and the second term accounts for the diamagnetic shift (black line in panel (d)).}
\label{fig:gX}
\end{center}
\end{figure*}

\subsection{ETB calculation of carrier $g$-factors and comparison with experiment}

To conclude on the role of quantum confinement, band mixing, and surface effects in 2D perovskites, we compare in Figure~\ref{fig:CUD} the energy dependences of the electron and hole $g$-factor in the 2D samples with the corresponding trends established for bulk lead halide perovskites~\cite{Kirstein_universal} and for CsPbI$_3$~\cite{Nestoklon_2023,Meliakov_2024_3} nanocrystals. In bulk lead halide perovskites, the evolution of electron and hole $g$-factors with band gap energy is governed primarily by band structure effects and follows a universal dependence across a wide range of compositions. In nanocrystals, a qualitatively similar dependence on the effective band gap has been reported. However, it is additionally influenced by the strong spatial confinement, which modifies the electronic structure and enhances the band mixing. As a result, deviations from the bulk universal trend are observed in the NCs. These deviations are well accounted for by empirical tight-binding (ETB) calculations for the electrons, see the closed circles and the dashed lines in Figures~\ref{fig:CUD}a,c (here the same data for NCs are shown in both panels, as in an ensemble of randomly oriented NCs the $g$-factor anisotropy is not pronounced). On the other hand, for holes the ETB theory predicts small deviations from the bulk dependence, while the experiments show larger changes (compare the brown circles and red dashed lines in Figures~\ref{fig:CUD}b,d).

The experimental results for the electron and hole $g$-factors in 2D and NCs shown in Figure~\ref{fig:CUD} are rather close to each other, showing a strong deviation from the bulk universal dependence. From that we conclude that quantum confinement provides the main contribution to the $g$-factor renormalization. However, the 2D and NC data demonstrate differences in various geometries, due to the fact that the carrier $g$-factors in 2D are anisotropic. 

For deeper insight, we perform ETB calculations of the electron and hole $g$-factors in 2D perovskites shown by the green lines in Figure~\ref{fig:CUD} as a function of exciton energy (details of the ETB calculations are in Supplementary Information \ref{sec:SI:calcul}). For that we calculate the carrier $g$-factors in thin platelets of the cubic Cs$_{N}$Pb$_{N+1}$I$_{3N+2}$ perovskite, following the methods developed in refs.~\cite{Nestoklon_2021,Kirstein_universal,Nestoklon_2023} as a function of effective band gap. The exciton binding energy calculated in ref.~\cite{Movilla2025} is used for correcting the effective band gap energy. The ETB theory predicts that the value of the in-plane $g_{\rm e, ab}$ (Voigt geometry) is expected to be strongly renormalized by quantum confinement, in comparison with the universal bulk dependence. In contrast, the out-of-plane $g_{\rm e, c}$ (Faraday geometry) is predicted to be only slightly changed.  One can see in Figure~\ref{fig:CUD}a that good quantitative agreement is achieved between theory and experiment for the electron $g$-factor in the Voigt geometry. For the Faraday geometry, only the qualitative trend agrees (see Figure~\ref{fig:CUD}c). It is clear that additional factors contribute to the electron $g$-factor anisotropy. We also refer the reader to the comparison between the experimental and calculated results shown in Figure~\ref{fig:MUD}c.
 
The clarification of their origins calls for further studies. We suggest that possible reasons can be related to the effect of the bulk $g$-factor anisotropy, which is well pronounced in MAPbI$_3$ crystals~\cite{Kirstein_universal} and/or to the additional effect of the lateral confinement and the Rashba effect, which is expected for PEA-based 2D perovskites~\cite{Zhai_2017}. 

For the holes in 2D perovskites, the ETB theory predicts only a small deviation from the universal bulk dependence, similar to the predictions for NCs, compare solid and dashed red lines in Figures~\ref{fig:CUD}b,d. The experiments on the 2D samples show a much stronger renormalization, in line with the experimental data for NCs. The underlying physical reasons for this behavior call for further studies.

\subsection{Layer dependence of exciton $g$-factor}

To determine the exciton $g$-factor and reconstruct its dependence on the number of layers, we measured the splitting of the exciton spin states in magnetic fields up to $B_\text{F} = 55$~T, applied in the Faraday geometry. Examples of circularly-polarized reflectivity spectra measured at $+55$~T and $-55$~T are shown in Figure~\ref{fig:gX}(a) for $n = 3$ and $n = 5$. A well-resolvable Zeeman splitting of the exciton resonances can be seen. 

The magnetic field-induced energy shift of the exciton spin states is shown in Figure~\ref{fig:gX}(b) for the sample with $n=3$. The dependence is nonlinear and can be described by the contributions from the Zeeman splitting and the diamagnetic shift:
\begin{equation}
\label{eq:DS}
E_\text{X}(B_\text{F}) - E_\text{X}(B_\text{F} = 0) = \pm \frac{1}{2}g_\text{X,c} \mu_B B_\text{F} + c_0 B_\text{F}^2,
\end{equation}
where $E_\text{X}(B_\text{F} = 0)$ is the exciton resonance energy at zero magnetic field, the second term corresponds to the Zeeman splitting, and the third term accounts for the quadratic diamagnetic shift with the coefficient $c_0$. The diamagnetic shift, exhibiting a quadratic dependence on the magnetic field, is presented in Figure~\ref{fig:gX}(d). From the corresponding fit we evaluate the diamagnetic coefficient: $c_{0,3} = 0.90 \pm 0.01\,\mu$eV/T$^{2}$. 

The Zeeman splitting, which is linear in magnetic field, is shown in Figure~\ref{fig:gX}(c).
It has no offset at zero magnetic field, and the slope of the linear dependence allows us to determine the $g$-factor. For $n = 3$, we obtain $g_\text{X,c} = +1.99 \pm 0.01$. Note that in this experiment the sign of the exciton $g$-factors can be determined. The stronger high energy shift of the $\sigma^+$ polarized resonance for positive magnetic fields shown in Figure~\ref{fig:gX}(a) allows us to conclude that the exciton $g$-factor has a positive sign. 

\begin{figure}[t]
\begin{center}
\includegraphics[width = 0.48\textwidth]{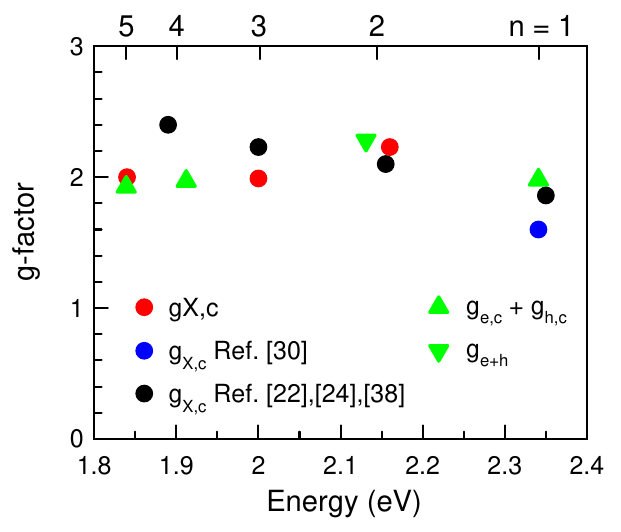}
\caption{Overview of the exciton $g$-factors in $(\mathrm{PEA})_{2}\mathrm{MA}_{n-1}\mathrm{Pb}_{n}\mathrm{I}_{3n+1}$ ($n = 1-5$). The red circles give the exciton $g$-factors determined from the Zeeman splitting in high magnetic fields at $T = 1.6$~K. The green symbols represent the sum of the electron and hole $g$-factors ($g_\text{e,c} + g_\text{h,c}$) and the $g$-factor determined from the combined spin-flip signal ($g_\text{e+h}$), measured in the Faraday geometry.  The blue and black circles correspond to the data reported in Refs.~\cite{Harkort_2023} and~\cite{Posmyk_2024,Dyksik_2021,Nikiforov_2026}, respectively. 
 }
\label{fig:X}
\end{center}
\end{figure}

The measured exciton $g$-factors for $n = 2,3,5$ are summarized in Figure~\ref{fig:X} (red circles) and in Table~\ref{Table_II}.
For completeness, we include the value of $g_{\text{X,c}}$ for $n = 1$ from our previous work~\cite{Harkort_2023} (blue circle). The exciton $g$-factors are in the range of $2-2.2$ for the samples with $n > 1$ and are nearly independent of the number of layers. The sum of the electron and hole $g$-factors ($g_{\text{e,c}} + g_{\text{h,c}}$), measured in the Faraday geometry for $n > 1$, is in good agreement with the exciton $g$-factor. However, for $n = 1$, we find that $g_{\text{e,c}} + g_{\text{h,c}} > g_{\text{X,c}}$, which can be attributed to the large exciton binding energy and Coulomb renormalization effects~\cite{Nikiforov_2026}.   

For comparison, we also include literature data from Refs.~\cite{Posmyk_2024,Dyksik_2021},  where the out-of-plane exciton $g$-factor was measured in reflectivity spectra (black symbols in the figure). These data are in good agreement with our results. 

\begin{table*}
\caption{Summary of the exciton $g$-factors measured from reflectivity and $g_{\rm e}+g_{\rm h}$ taken from Table~\ref{Table_I} in (PEA)$_2$MA$_{n-1}$Pb$_n$I$_{3n+1}$ samples, evaluated from SFRS and TRKR experiments. The $g$-factor measurement accuracy is $\pm$0.1 for the reflectivity measurements.}
\centering
\begin{tabular}{p{0.7cm} p{1.8cm} p{1.2cm}  p{2cm} p{2cm} }
%\begin{tabular}{rrrrr}
\hline
%& & SFRS & SFRS & SFRS & SFRS & TRKR & TRKR\\
$n$ & $E_\mathrm{exc}$ & $g_\mathrm{X,c}$ & $g_{\rm e,c}+g_{\rm h,c}$ & $g_{\rm e,ab}+g_{\rm h,ab}$ \\
\hline
1 & 2.341~eV & $+1.6$  & 1.98 &  1.99 \\
2 & 2.129~eV & $+2.0$  &  & 2.28  \\
3 & 2.010~eV & $+1.9$  &  &   \\
4 & 1.908~eV & -  & 1.97 & 2.21  \\
5 & 1.837~eV &  $+1.9$ & 1.926 & 2.80  \\
\hline
\end{tabular}
\label{Table_II}
\end{table*}

\section{Conclusions}

We have used magneto-optical techniques to measure the Zeeman splittings of the electrons and holes as well as the excitons in (PEA)$_2$MA$_{n-1}$Pb$_n$I$_{3n+1}$ 2D perovskites with $n=1,...,8$ and evaluate their Landé $g$-factors from these data. With reducing $n$, i.e., increasing  the quantum confinement, a strong renormalization of the electron and hole $g$-factors is found, which is comparable to the changes previously reported in perovskite nanocrystals. These results are in good qualitative agreement with empirical tight-binding calculations. In 2D perovskites, similar to bulk and nanocrystals, the strong changes of the electron and hole $g$-factors approximately compensate each other, resulting in nearly constant values of the bright exciton $g$-factor for samples with different $n$.

\section{Materials and Methods}

\textbf{Samples.}
The experiments are performed on Ruddlesden-Popper type 2D perovskites of the composition PEA$_2$MA$_{n-1}$Pb$_n$I$_{3n+1}$, deposited on glass substrates. The samples contain a mixture of different numbers of inorganic layers of the lead halide octahedra ($n = 1, ..., 8$). 

\textbf{(PEA)$_2$PbI$_{4}$ thin films with $n = 1$.}  For the synthesis of these thin films, (PEA)I and PbI$_2$ were dissolved in N,N-dimethylformamide in the stoichiometric molar ratio to prepare a solution of 8.3 wt. \% concentration of (PEA)$_2$PbI$_{4}$. Inside a nitrogen-filled glovebox, the solution was spin-coated on glass substrates for 5 s at 400 rpm, followed by 30 s at 3000 rpm. The samples were then annealed on a hot plate at $100^\circ$C for 10 minutes.

\textbf{(PEA)$_2$MA$_{n-1}$Pb$_{n}$I$_{3n+1}$ single crystals with $n > 1$.} The synthesis of these single crystals is based on the cooling method described
in ref.~\cite{Peng_2017}. The precursors, including lead oxide, methylammonium iodide, and phenethylamine, were dissolved in hydriodic acid at the specific molar ratios and maintained at approximately $110^\circ$C for 4 hours. The solutions were then gradually cooled to room temperature at a controlled rate of $1^\circ$C per hour. The molar ratios of lead oxide, methylammonium iodide, and phenethylamine were 10/24/1 mmol, respectively. The obtained crystals were filtered under vacuum and then dried with diethylether.

\textbf{Photoluminescence and absorption measurements.} The samples are immersed in superfluid liquid helium ($T=1.6$~K). For the photoluminescence measurements, the samples are excited with a continuous wave laser, while for measuring the absorption a halogen lamp is used. The spectra are measured with an 0.5~m spectrometer equipped with a charge-coupled-device (CCD) camera. 

\textbf{Time-resolved Kerr rotation (TRKR).} The coherent spin dynamics are studied by a pump-probe time-resolved Kerr rotation technique~\cite{Awschalom2002,Yakovlev_ch6_2008}. Pump and probe light have the same photon energy. In order to provide resonant excitation of the exciton states in different samples we use an optical parametric oscillator (OPO), pumped by a pulsed titanium-sapphire (Ti:Sa) laser with a pulse duration of 1.5~ps and a spectral width of about $1$~nm (1.5~meV). The laser  repetition rate is 76~MHz, which corresponds to a repetition period of $T_\text{R}=13.2$~ns. the laser photon energy is tunable across the spectral range from $1.76-2.40$~eV. The probe pulses are sent pver a mechanical delay line and thereby become time-delayed with respect to the pump pulses. The pump helicity is modulated between $\sigma^+$ and $\sigma^-$ circular polarization at the frequency of 50~kHz by a photo-elastic modulator (PEM). The probe polarization is linear, and its amplitude is modulated at the frequency of 84~kHz by a PEM. The polarization of the reflected probe beam is analyzed via balanced photodiodes and a lock-in amplifier. The samples are immersed in superfluid liquid helium ($T=1.6$~K). Magnetic fields up to 3~T are applied in the Voigt geometry, i.e. perpendicular to the laser wave vector direction and the induced carrier spin polarization.

The Kerr rotation signal shows decaying oscillations in time in the transverse magnetic field, due to the Larmor precession combined with spin relaxation of the electrons and holes. In perovskite semiconductors, both the electron and hole spin coherences are observed in the Kerr rotation signal. The signal can be described as the sum of two decaying cosine functions: 
\begin{equation}
\begin{aligned}
A_{\rm KR} = &S_{\rm e} \cos (\omega_{\rm L, e} t) \exp(-t/T^*_{\rm 2,e}) + \\ &S_{\rm h} \cos (\omega_{\rm L, h} t) \exp(-t/T^*_{\rm 2,h}) \,.
\label{eqn:KR}
\end{aligned}
\end{equation}
 Here $S_{\rm e(h)}$ are the electron and hole spin polarizations right after the pump action. $T^*_{\rm 2,e(h)}$ are the carrier spin dephasing times. The Larmor precession frequencies $\omega_{\rm L, e(h)}$ are used to evaluate the $g$-factors according to $|g_{\rm e(h)}|=  \hbar \omega_{\rm L, e(h)}/ (\mu_{\rm B} B)$. Note that the TRKR signal does not provide direct information about the $g$-factor sign. 

\textbf{Spin-flip Raman scattering (SFRS).} 
The Zeeman splitting of the resident carrier spin sublevels can be measured directly by the SFRS technique~\cite{hafele_chapter_1991,Debus_2013,Rodina_2022,Kalitukha_2026}. The light scattered from the sample has a spectral shift (Raman shift) from the laser photon energy, due to the phonon assisted spin-flip of the carriers. 
%For the semiconductors, Raman shift is about 1~meV at magnetic field of 10~T. 
The samples are immersed in superfluid liquid helium ($T=1.6$~K). The SFRS signal is excited by a laser with photon energy coinciding with the exciton resonance, therefore the laser is tunable from 1.75~eV to 3.1~eV (Syrah laser system based on a Ti:Sapphire laser with the spectral range extended by a frequency doubling unit and by mixing with the light of a fiber laser emitting at 1550~nm). The laser power density of 0.15~Wcm$^{-2}$ is used. The SFRS signal is measured in the backscattering geometry. A Jobin-Yvon U1000 double monochromator with 1 meter focal length is used to detect the SFRS spectra with the high spectral resolution of 0.2~cm$^{-1}$ (0.024~meV). The signal is detected by a cooled GaAs photomultiplier and conventional photon counting electronics. The spectrally narrow laser and the high spectral resolution of the spectrometer combined with efficient suppression of the scattered laser light allow us to detect Raman shifts from 0.1 to 3~meV. The SFRS spectra are measured for cross-polarized ($\sigma^- / \sigma^+$) circular polarizations of excitation ($\sigma^-$) and detection ($\sigma^+$) to suppress the scattered laser light. The SFRS signal in the studied samples is polarization-independent. therefore, this polarization configuration was used for both the Faraday and Voigt geometries. In the Faraday geometry, the magnetic field is parallel to the light wave vector ${\bf k}$ and to the $c$-axis of the crystal ($\textbf{B}_\text{F} \parallel \textbf{k}$, $\textbf{B}_\text{F} \parallel c$). In the Voigt geometry, the field is perpendicular to these vectors ($\textbf{B}_\text{V} \perp \textbf{k}$, $\textbf{B}_\text{V} \parallel (a,b)$).   

\textbf{Reflectivity in pulsed magnetic fields.} 
The Zeeman splitting of excitons is measured in high magnetic fields up to $\pm55$~T, using a capacitor-driven pulsed magnet (rise time $t=32$\,ms).  The sample is mounted on a custom fiber-coupled probe in a helium bath cryostat with a long tail extending into the bore of the magnet. The experiments are performed at $T = 1.6$~K, with the sample immersed in superfluid helium. Broadband white light from a halogen lamp is coupled down a 100~$\mu$m-diameter multimode optical fiber, and light reflected from the sample is collected by a 600~$\mu$m-diameter fiber. The light wavevector $\bf k$ is perpendicular to the sample surface and parallel to $\bf B$ (Faraday geometry). A thin-film brodband circular polarizer between the sample and the collection fiber is used for polarization-resolved measurements. Reflectivity spectra are acquired continuously throughout the magnet pulse using a fast charge-coupled-device camera combined with a 0.3 m spectrometer.  Typical acquisition times range from 1.5 to 4.5 ms, depending on the sample. To switch between the $\sigma^-$ and $\sigma^+$ circular polarizations, the direction of the magnetic field is switched. Further technical details can be found in ref.~\cite{Stier_2016}.

\subsection*{Data Availability Statement}
The data presented in this paper are available from the corresponding authors upon reasonable request.

\subsection*{Supporting Information}
Additional spin-flip Raman spectra, $g$-factor evaluation, and $g$-factor anisotropy for samples with $n = 4, 5$, and 7. Details of the ETB calculations of the electron and hole $g$-factors.

\subsection*{Acknowledgements}
%The authors are thankful to \aN{XXX} and \aN{XXX} for fruitful discussions.  
N.E.K. acknowledges the support of the Deutsche Forschungsgemeinschaft (project KO 7298/1-1, no. 552699366). D.R.Y, D.K., and C.H. acknowledge the financial support by the Deutsche Forschungsgemeinschaft via the SPP 2196 Priority Program (project YA 65/28-1, no. 527080192). M.O.N. acknowledges the Deutsche Forschungsgemeinschaft for support via the SPP 2196 Priority Program  (project no. 506623857). The work at ETH Z\"urich (O.H., O.F.D., and M.V.K.) was financially supported by the Swiss National Science Foundation (grant agreement 200020E 217589) through the DFG-SNSF bilateral program and by the ETH Z\"urich through the ETH+ Project Syn-MatLab. The National High Magnetic Field Laboratory is supported by the US National Science Foundation DMR-2128556, the US Department of Energy, and the State of Florida.

\textbf{ORCID}\\
Nataliia E. Kopteva:  0000-0003-0865-0393 \\ % natalia.kopteva@tu-dortmund.de
Dmitri R. Yakovlev:   0000-0001-7349-2745 \\ % dmitri.yakovlev@tu-dortmund.de
Mikhail O. Nestoklon: 0000-0002-0454-342X \\ % nestoklon@gmail.com
Carolin Harkort:      0000-0003-1975-9773 \\ % carolin.harkort@tu-dortmund.de
Evgeny A. Zhukov:     0000-0003-0695-0093 \\% evgeny.zhukov@tu-dortmund.de
Dennis Kudlacik:      0000-0001-5473-8383 \\ % dennis.kudlacik@gmail.com
Erik Kirstein:        0000-0002-2549-2115 \\  %erik.kirstein@udo.edu
Scott A. Crooker:     0000-0001-7553-4718 \\  %crooker@lanl.gov
Oleh Hordiichuk:      0000-0001-7679-4423 \\  % holeh@student.ethz.ch
Ole F. Dressler: 0000-0002-1068-9898       \\ % odressler@ethz.ch
Maksym V. Kovalenko:  0000-0002-6396-8938 \\ % mvkovalenko@ethz.ch
Manfred Bayer:        0000-0002-0893-5949 \\ % manfred.bayer@tu-dortmund.de

\clearpage
\newpage
\onecolumngrid
\begin{center}
  \textbf{{\Large Supporting Information\\ Land\'e $g$-factor of electrons and holes in two-dimensional Ruddlesden-Popper lead halide perovskites }}\\
\vspace{5mm}
{Nataliia E. Kopteva$^{1}$, Dmitri~R.~Yakovlev$^{1}$, Mikhail~O.~Nestoklon$^{1}$, Carolin~Harkort$^{1}$, Evgeny~A.~Zhukov$^{1}$, Dennis Kudlacik$^{1}$, Erik Kirstein$^{1,2}$, Scott A. Crooker$^{2}$,  Oleh~Hordiichuk$^{3,4}$, Ole Dressler$^{3,4}$,  Maksym~V.~Kovalenko$^{3,4}$, and Manfred~Bayer$^{1,5}$}\\
  
\vspace{5mm}  
  {\textit{$^1$Experimentelle Physik 2, Technische Universit\"at Dortmund, 44221 Dortmund, Germany\\ 
  {$^{2}$National High Magnetic Field Laboratory, Los Alamos National Lab,
Los Alamos, NM 87545, USA} \\
  $^{3}$Laboratory of Inorganic Chemistry, Department of Chemistry and Applied Biosciences,  ETH Z\"{u}rich, CH-8093 Z\"{u}rich, Switzerland\\ 
  $^{4}$Laboratory for Thin Films and Photovoltaics, Empa-Swiss Federal Laboratories for Materials Science and Technology, CH-8600 D\"{u}bendorf, Switzerland\\ 
  $^{5}$Research Center FEMS, Technische Universit\"at Dortmund, 44227 Dortmund, Germany}\par} 
\end{center}

\setcounter{equation}{0}
\setcounter{figure}{0}
%\setcountevener{figure}{0}
\setcounter{table}{0}
\setcounter{page}{1}
\setcounter{section}{0}
\makeatletter
\renewcommand{\thepage}{S\arabic{page}}
\renewcommand{\theequation}{S\arabic{equation}}
\renewcommand{\thefigure}{S\arabic{figure}}
\renewcommand{\thetable}{S\arabic{table}}
\renewcommand{\thesection}{S\arabic{section}}
\renewcommand{\bibnumfmt}[1]{[S#1]}

\makeatletter

\makeatother

\section{SFRS data for samples with $\bf n = 4,5,7$}
\label{sec:addSFRS}

\begin{figure*}[tbh]
\begin{center}
\includegraphics[width = \textwidth]{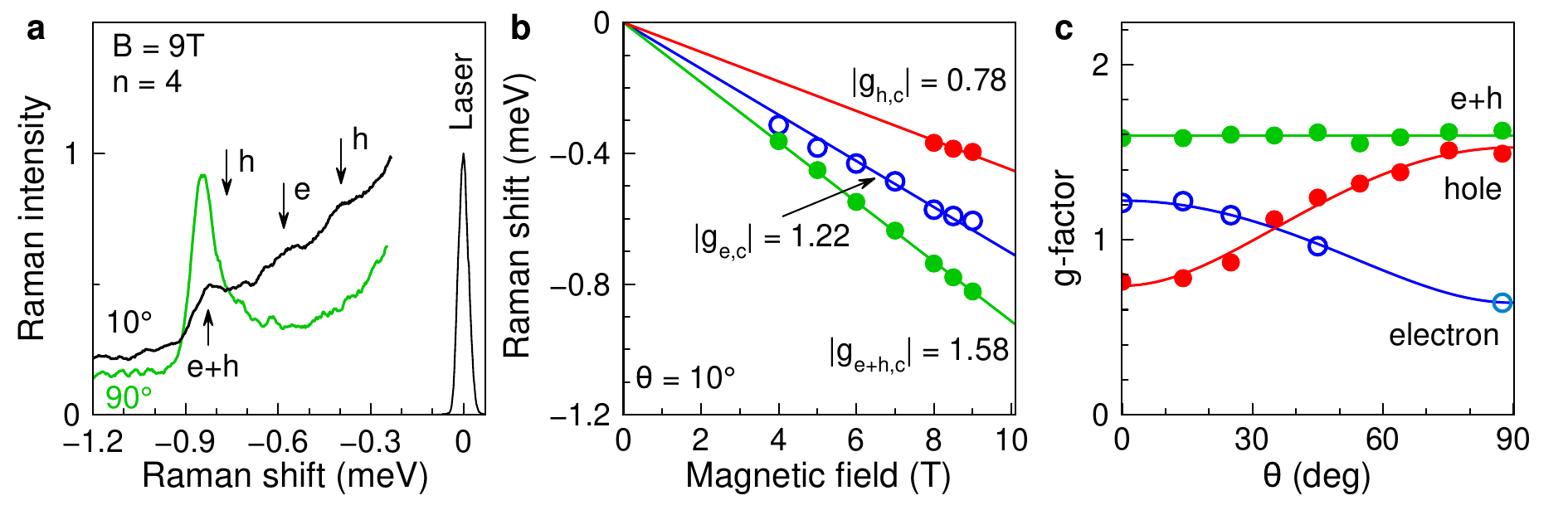}
\caption{Spin-flip Raman scattering of electrons and holes in (PEA)$_2$MA$_3$Pb$_4$I$_{13}$ ($n = 4$) at $T =1.6$\,K. 
(a) SFRS spectra measured at $B = 9$~T applied in the Voigt geometry ($\theta = 90\degree$, green) and tilted geometry ($\theta = 10\degree$, black) in the $n = 4$ sample. 
$E_\text{exc} = 1.913$\,eV and $P = 0.15$\,W/cm$^2$. The spin-flip transitions of electrons (e), holes (h), and combined $e+h$ states are indicated by the vertical arrows.
(b) Raman shifts of the holes (red symbols), electrons (blue symbols), and combined $e+h$ (green symbols) as function of the magnetic field applied in tilted geometry ($\theta = 10\degree$). The solid lines are fits using eq~\eqref{eqn:EZ}, yielding $g_\text{e,c} = 1.22$, $g_\text{h,c} = 0.78$, and $g_\text{e+h,c} = 1.58$. 
(c) Angular dependence of the hole (red symbols), electron (blue symbols), and combined $e+h$ (green symbols) $g$-factors. The solid lines are fits with eq~\eqref{eqn:ani_gf}.
}
\label{fig:SI_n3}
\end{center}
\end{figure*}

\begin{figure*}[t!]
\begin{center}
\includegraphics[width = \textwidth]{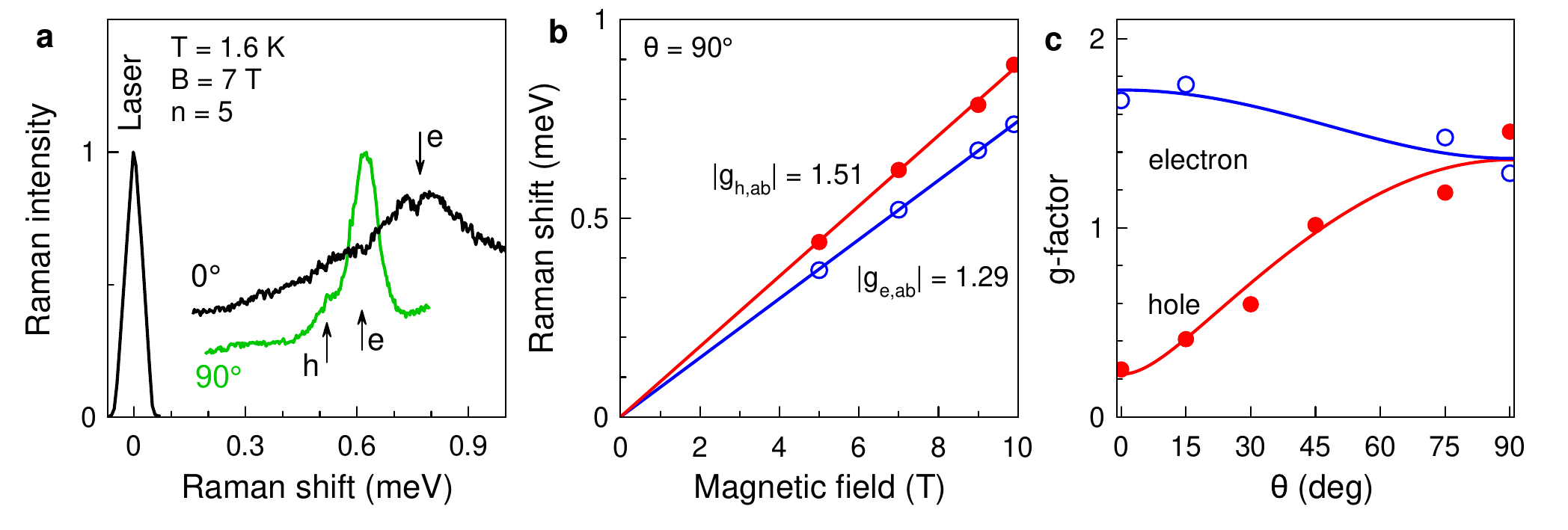}
\caption{Spin-flip Raman scattering of electrons and holes in (PEA)$_2$MA$_4$Pb$_5$I$_{16}$  ($n = 5$) at $T =1.6$\,K. 
(a) SFRS spectra measured at $B = 7$~T applied in the Voigt geometry ($\theta = 90\degree$, green) and Faraday geometry ($\theta = 0\degree$, black). 
$E_\text{exc} = 1.839$\,eV and $P = 0.15$\,W/cm$^2$. The spin-flip transitions of the electrons (e) and holes (h) are indicated by the vertical arrows.
(b) Raman shifts of the holes (red symbols) and electrons (blue symbols) as function of the magnetic field applied in tilted geometry ($\theta = 90\degree$). The solid lines are fits using eq~\eqref{eqn:EZ}, yielding $g_\text{e,ab} = 1.29$ and $g_\text{h,ab} = 1.51$. 
(c) Angular dependence of the hole (red symbols) and electron (blue symbols) $g$-factors. The solid lines are fits with eq~\eqref{eqn:ani_gf}.}
\label{fig:SI_n4}
\end{center}
\end{figure*}

\begin{figure*}[t!]
\begin{center}
\includegraphics[width = \textwidth]{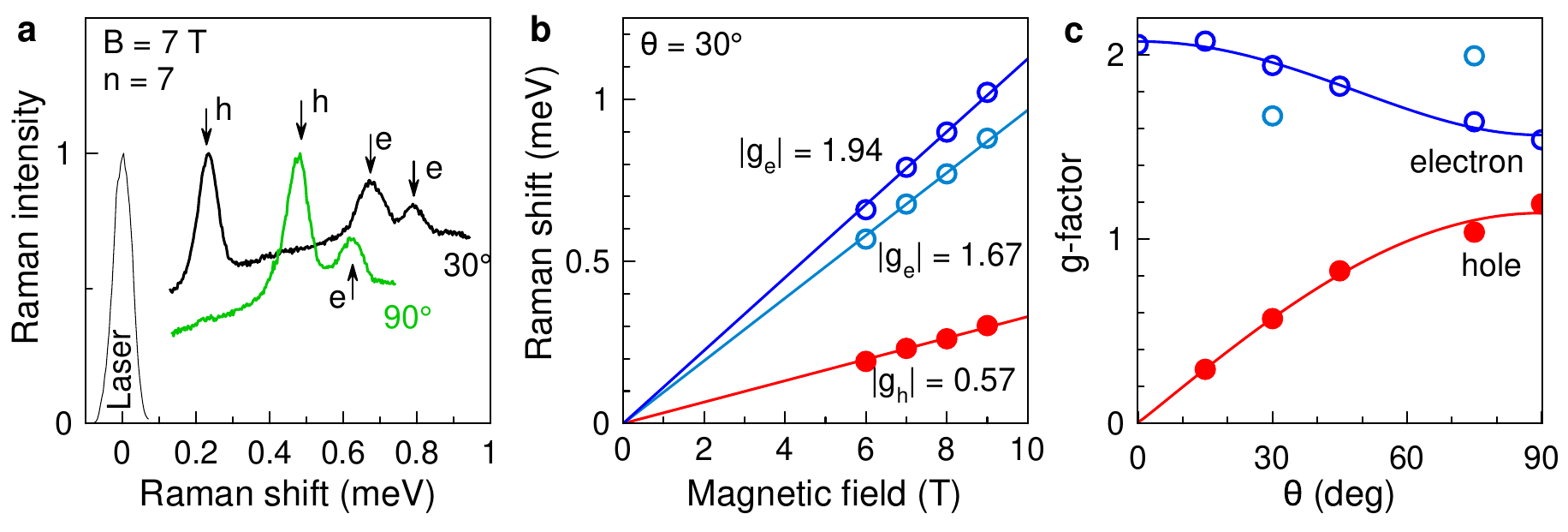}
\caption{Spin-flip Raman scattering of electrons and holes in (PEA)$_2$MA$_6$Pb$_7$I$_{22}$  ($n = 7$) at $T =1.6$\,K. 
(a) SFRS spectra measured at $B = 7$~T applied in the Voigt geometry ($\theta = 90\degree$, green) and tilted geometry ($\theta = 30\degree$, black). 
$E_\text{exc} = 1.798$\,eV and $P = 0.15$\,W/cm$^2$. The spin-flip transitions of the electrons (e) and holes (h) are indicated by the vertical arrows.
(b) Raman shifts of the holes (red symbols), and electrons (blue and light blue symbols) as function of magnetic field applied in tilted geometry ($\theta = 30\degree$). The solid lines are fits using eq~\eqref{eqn:EZ}, yielding $g_\text{e} = 1.94$, $g_\text{e} = 1.67$, and $g_\text{h} = 0.57$. 
(c) Angular dependence of the hole (red symbols) and electron (blue and light blue symbols) $g$-factors. The solid lines are fits with eq~\eqref{eqn:ani_gf}.}
\label{fig:SI_n6}
\end{center}
\end{figure*}

Figures~\ref{fig:SI_n3}-\ref{fig:SI_n6} present spin-flip Raman scattering (SFRS) data for the Ruddlesden–Popper perovskites (PEA)$_2$MA$_{n-1}$Pb$_n$I$_{3n+1}$ with inorganic layer numbers of $n = 4$, $n = 5$, and $n = 7$, respectively. The figures are organized identical to facilitate their direct comparison. For each composition, panel (a) shows SFRS spectra measured at $T = 1.6$~K for different magnetic field orientations, allowing identification of the electron and hole spin-flip transitions. Panel (b) displays the magnetic field dependence of the corresponding Raman shifts, from which the in-plane electron and hole $g$-factors are evaluated using eq~\eqref{eqn:EZ}. Panel (c) presents the angular dependence of the $g$-factors, which can be described by the anisotropic $g$-factor model given by eq~\eqref{eqn:ani_gf}.

All together, Figures~\ref{fig:SI_n3}-\ref{fig:SI_n6} demonstrate that spin-flip Raman scattering provides access to the electron and hole $g$-factors across the 2D sample set with different $n$. They confirm the systematic evolution of the $g$-factor values and anisotropied with the number of inorganic layers, as discussed in the main text.

\clearpage

\section{Exciton parameters}
\label{sec:gUD_Voigt}

\begin{table*}[hbt]
\centering
\caption{Exciton parameters in the (PEA)$_2$MA$_{n-1}$Pb$_n$I$_{3n+1}$ samples measured at $T = 1.6$\,K. The exciton binding energies were calculated in Ref.~\cite{Movilla_2021} (calculations with Bajaj potential, Fig.~2d).
}
\begin{tabular}{|c|c|c|c|c|c|c|c|c|}
\hline
 & $n = 1$ & $n = 2$ & $n = 3$ &$n = 4$ & $n = 5$ & $n = 6$ & $n = 7$ & $n = 8$\\ \hline
$E_\text{X}$ (eV) & 2.341 & 2.129 & 2.010 &1.908 & 1.837 & 1.813 & 1.795 & 1.762\\ \hline
FWHM (meV) & 6.6 & 11 & 28 & 9 & - &12 & 9.2 & 13.2\\ \hline
Stokes shift (meV) & 0 & 12 & 6 & 12 & 8 & 1 & 1 & 1\\ \hline  
Binding energy (meV) \cite{Movilla_2021}& 271 & 154 & 111 & 88.5  & 74.7 & 64.5 & 57.8 & 51.4\\ \hline 
\end{tabular}
\label{tab:exciton}
\end{table*}

\section{Calculation of electron and hole $g$-factors in 2D perovskites}
\label{sec:SI:calcul}

\begin{figure*}[hbt]
\begin{center}
\includegraphics[width = 0.6\textwidth]{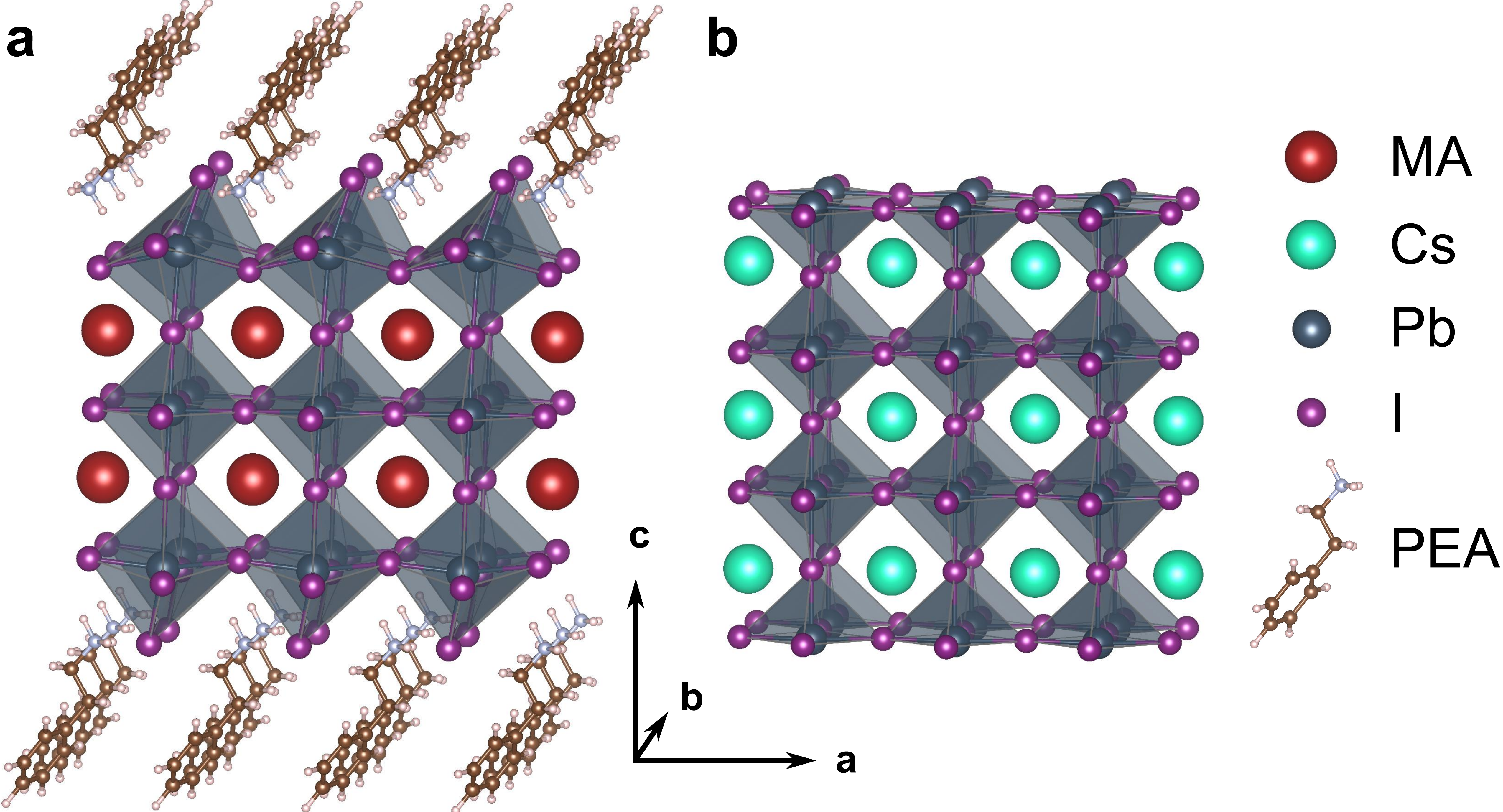}
\caption{(a) Scheme of atomic arrangement in (PEA)$_2$MA$_{2}$Pb$_3$I$_{10}$ ($n=3$). (b) Scheme of the structure used in the ETB calculations of the $g$-factors for $N=3$: Cs$_{3}$Pb$_{4}$I$_{11}$. The crystallographic axes are shown by the arrows.}
\label{fig:surfaces}
\end{center}
\end{figure*}

To analyze the $g$-factors of electrons and holes in 2D perovskites including their dependence on the number of layers, we performed empirical tight-binding (ETB) calculations. We use the original code with straightforward implementation of the Slater-Koster scheme \cite{Slater1954} for the construction of the sparse ETB Hamiltonian using the parameters given in table~S1 of ref.~\cite{Nestoklon_2023}. Then, a few selected eigenvalues of the ETB Hamiltonian near the band gap are found using the self-written implementation of the Thick Restarted Lanczos algorithm \cite{TRLan}.  In ref.~\cite{Nestoklon_2023} this approach was successfully used for calculation of the $g$-factors of charge carriers in cubic perovskite nanocrystals (NCs). We compute the 2D structures with formula Cs$_{N}$Pb$_{N+1}$I$_{3N+2}$,  
which have fixed a number of monolayers $N$ in the [001] direction (i.e., along the c-axis) and square $M\times M$ unit cells in the (001) plane (i.e., in the ab-plane). The calculated structures are different from the experimental ones in the surface termination (see Fig.~\ref{fig:surfaces}): exact modeling of the actual surface termination needs accurate accounting of the passivation of the I termination by the PEA layers with surface relaxation, charge redistribution, etc. The role of the surface passivation is expected to have a minor effect on the $g$-factors for structures with large $N$ and requires a separate investigation for the case of small $N = 2,3$. As demonstrated in ref.~\cite{Nestoklon_2021}, the PbI$_2$ termination (crystal is ``cut'' through the (001) plane containing the Pb atoms) does not produce surface states without explicit passivation. In contrast, for the halide termination proper passivation is necessary, which is beyond the capabilities of our ETB model. 

\begin{figure*}[hbt]
\begin{center}
\includegraphics[width = 0.35\textwidth,page=4]{FigsTheory.pdf}%\hspace{0.000005\textwidth}%
\includegraphics[width = 0.65\textwidth,page=1]{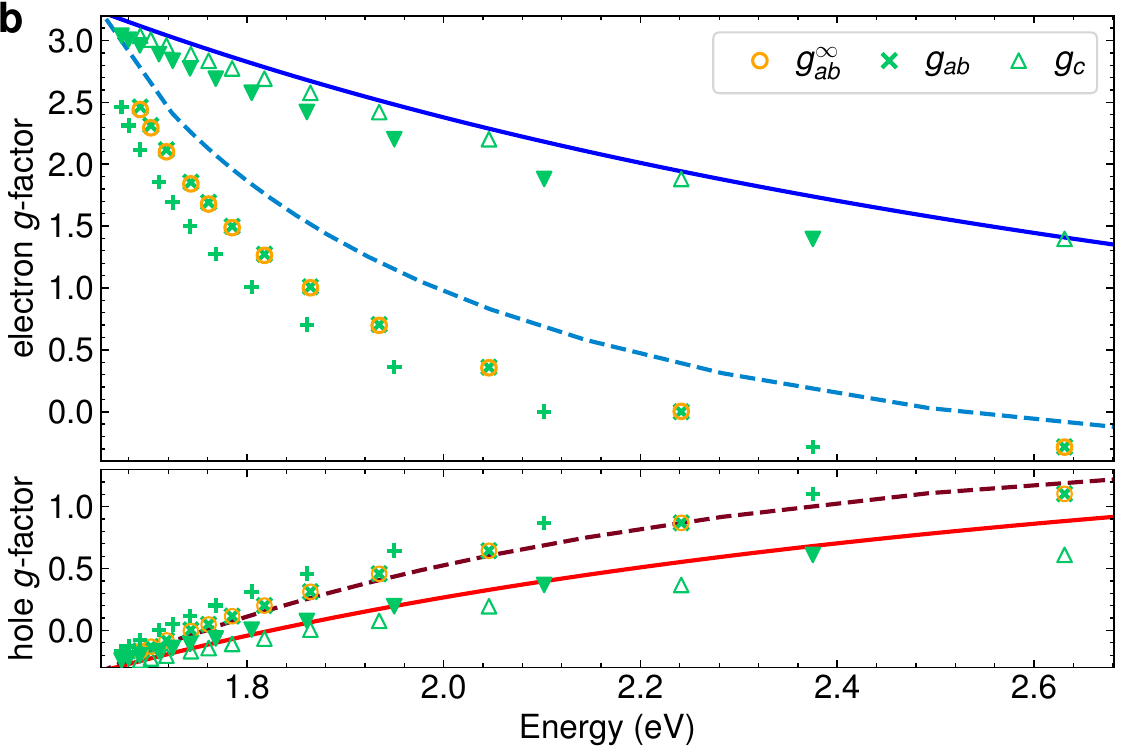}
\caption{(a) The dots show the effective band gap energy, i.e., the difference between the electron and hole energies in the structure Cs$_{N}$Pb$_{N+1}$I$_{3N+1}$ with $N$ monolayers calculated in ETB. By the ``+'' symbols we show the ETB calculations corrected for the exciton binding energy calculated in ref.~\cite{Movilla_2021}. (b) The open triangles and crosses show the electron and hole $g$-factors as function of the effective band gap energy calculated using the ETB method, see text. The crosses show the $g$-factors calculated for the in-plane magnetic field ${\bf B}\| [100]$ (Voigt geometry) and the triangles for the in-plane field ${\bf B}\| [001]$ (Faraday geometry). The orange circles show the in-plane $g$-factors calculated for infinite structures. The dashed lines reproduce the results of the ETB calculations for CsPbI$_3$ nanocrystals from ref.~\cite{Nestoklon_2023}. The solid lines are the results for the universal bulk dependence from ref.~\cite{Kirstein_universal}. The closed triangles and ``+'' symbols show the results corrected for the exciton binding energy from Ref.~\cite{Movilla_2021}
}
\label{fig:gETB1}
\end{center}
\end{figure*}

To avoid the well-known problem of incompatibility between periodic boundary conditions and magnetic field~\cite{Graf_1995,Hofstadter_1976} for the field applied in the Faraday geometry (along [001]), we calculated the values of $g_{\rm c}$ in structures of finite size. Using the ETB method with the Peierls substitution following ref.~\cite{Graf_1995}, the electron and hole energies in a small finite magnetic field directed along [001] and [100] were calculated and from the splitting of the two spin states, which are degenerate at zero magnetic field, the corresponding $g$-factors were evaluated: $g^{MN}_{\rm e, c}$, $g^{MN}_{\rm e, ab}$,  $g^{MN}_{\rm h, c}$, and $g^{MN}_{\rm h, ab}$. The $g$-factors in infinite structures then were obtained as the limits of the $g$-factors for a finite structure with $M \to \infty$. The limit is found by approximating the $g^{MN}_{\rm e, ab}$ at large $M$ by a linear function of energy. To validate this approach and check the convergence of the $g$-factors as function of $M$, we also calculate the $g_{\rm ab}^{\infty N}$: the $g$-factor for the field along the [100] direction (Voigt geometry) in infinite perovskite platelets. In this case, in contrast to a magnetic field along the [001] direction (Faraday geometry), the vector potential does not contradict the periodic boundary conditions, so that such calculations are straightforward. Then, one may easily check the convergence of the procedure and confirm that the limit of the $g$-factors in finite structures gives the correct value. We demonstrate that in fig.~\ref{fig:gETB1}b, where $g_{\rm ab}$ is calculated as the limit $g_{\rm ab}=\lim_{M\to \infty} g_{\rm e/h,ab}^{MN}$ (green crosses), while $g_{\rm ab}^{\infty}$ (orange circles) is calculated for the infinite 2D structure. One can see that these data sets coincide with each other.

We note that in the ETB calculations a number of assumptions is made: First, CsPbI$_3$ material is calculated and not MAPbI$_3$. Second, in the calculations we use PbI$_2$ surface termination and not PEA passivation. As a result, the unit formula used for the ETB calculations is Cs$_{N}$Pb$_{N+1}$I$_{3N+2}$, and not the experimental one of (PEA)$_2$MA$_{n-1}$Pb$_n$I$_{3n+1}$. Even though the structures have the same width, the boundary conditions are different. Finally, in the ETB calculations excitonic effects, especially important for the structures with small $n$, are neglected.
For better comparison with the experimental data, we correct the single particle band gap, calculated in ETB as difference of the energies of electron and hole in the structure, for the exciton binding energy calculated in ref.~\cite{Movilla_2021}. The corrected results are shown in Figure~\ref{fig:gETB1}. In the main text, Figure~\ref{fig:CUD}, the $g$-factors are shown as function of exciton energy.

The results of our calculations of the $g$-factors in 2D structures are shown in Figure~\ref{fig:gETB1}b by the symbols (see caption). For comparison, we also show by the dashed lines the results of the $g$-factors for cubic CsPbI$_3$ nanocrystals calculated in ref.~\cite{Nestoklon_2023} and by the solid lines we show the universal bulk dependence from ref.~\cite{Kirstein_universal}. Qualitatively, the results for the electron states may be understood taking into account the analysis for perovskite nanocrystals: the quantum confinement leads to mixing of the electron states with the higher lying spin-split electron band, which results in a renormalization of the $g$-factor, as compared with the universal bulk dependence. Note that in 2D samples this renormalization is anisotropic. For a magnetic field parallel to the $c$-axis (Faraday geometry) the renormalization is negligible,  while for an in-plain magnetic field (Voigt geometry) the renormalization is somewhat larger than for NCs for the same quantum confinement energy. In the latter case the renormalization is larger because in 2D structures the quantum confinement occurs only in one direction, while for NCs the electron is confined in all three directions. As a result, for the same confinement energy in 2D and in NCs, the confinement length is smaller in 2D so that the corresponding band mixing and resulting $g$-factor renormalization is larger. 

Next, we note that the renormalization of the electron $g$-factors by quantum confinement, which leads to the large anisotropy of the $g$-factor in 2D platelets, demonstrated in Figure~\ref{fig:gETB1}(b), is significantly modified by the lateral, i.e., the in-plane quantum confinement. The electron $g$-factor anisotropy is noticeably reduced for electrons localized in the lateral direction. 

\begin{figure*}[t]
\begin{center}
\includegraphics[width = 0.49\textwidth,page=2]{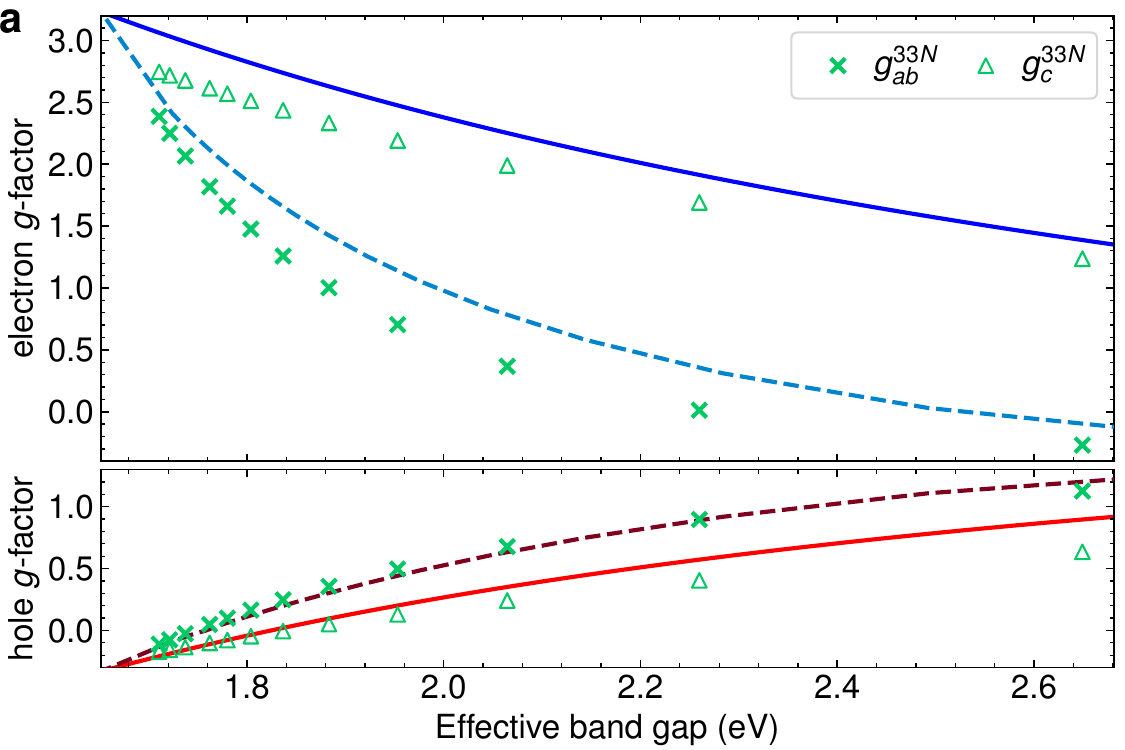}
\includegraphics[width = 0.49\textwidth,page=3]{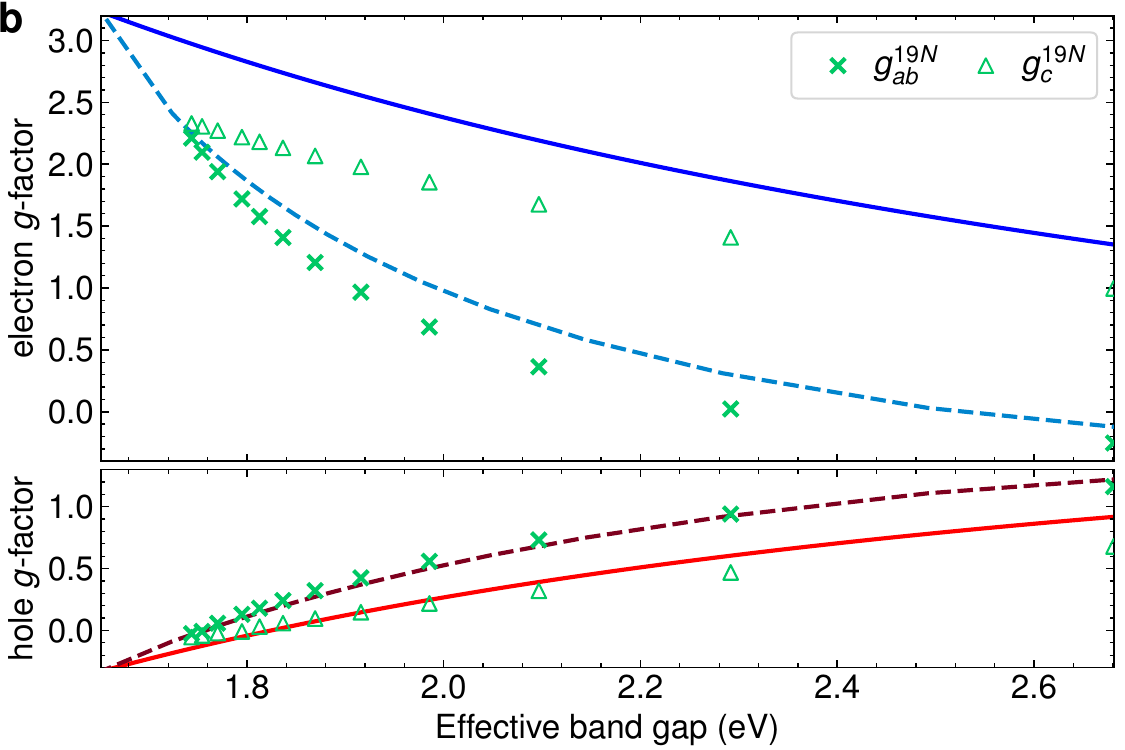}
\caption{Calculations of the electron and hole $g$-factors in 2D  Cs$_{N}$Pb$_{N+1}$I$_{3N+2}$ structures with finite in-plane localization of $M=33$ (a) and $M=19$ (b). The calculation results are shown by the symbols. The solid lines give the universal bulk dependences from ref.~\cite{Kirstein_universal}. The dashed lines give the ETB calculations for CsPbI$_3$ nanocrystals from ref.~\cite{Nestoklon_2023}. Note that here the exciton binding energy is not taken into account.
 }
\label{fig:gETB2}
\end{center}
\end{figure*}

To illustrate this behavior, in Figure~\ref{fig:gETB2} we show the $g$-factors calculated for cuboid quantum dots containing $M\times M \times N$ elementary cells with $M=33$ and $M=19$. We note that these values are large as compared with $n$ and correspond to the lateral localization of about $15$~nm and $10$~nm, respectively. They also exceed the exciton size in bulk material. For these relatively large structures the anisotropy of the electron $g$-factor is noticeably decreased for all thicknesses. The decrease is strong for larger $N$ (i.e., smaller confinement energies), while for small $N$ (i.e., large confinement energies) the effect of the finite $M$ is minor. This result demonstrates the importance of taking into account the in-plane localization in the analysis.

The calculations predict a similar behavior for the hole $g$-factors: In the infinite structure, the calculations give $g_{\rm h,ab}$ values close to the values for isotropic nanocrystals, while $g_{\rm h,c}$ is predicted to follow the universal bulk dependence. Taking into account the lateral confinement in 2D materials, the anisotropy is expected to be reduced, even though, for the holes both renormalization and anisotropy are expected to be much smaller than for the electrons. However, as discussed in ref.~\cite{Meliakov_2024_3}, our ETB calculations underestimate the renormalization of the hole $g$-factors as compared with the experimental data, see Fig.~\ref{fig:CUD}.

To conclude, the empirical tight-binding calculations of the shape-induced $g$-factor anisotropy predict that the anisotropy is large for thin platelets and reduces for thicker 2D structures. The origin of the anisotropy of the electron $g$-factor is analogous to the renormalization of the $g$-factor in nanocrystals and stems from the mixing with the second conduction band. As a result, $g_c$ is close to the universal bulk dependence \cite{Kirstein_universal}, while $g_{ab}$ is close to the $g$-factor in nanocrystals \cite{Nestoklon_2023}. For larger structures both tend to the bulk $g$-factor values. A similar behavior is expected for the hole $g$-factors. In addition, the calculations demonstrate that in the 2D structures with an intermediate number of layers the $g$-factor anisotropies depend sensitively to the lateral confinement, which reduces the anisotropy.

\end{document}